\documentclass[lettersize,journal]{IEEEtran}
\usepackage{amsmath,amsfonts}
\usepackage{mathrsfs}
\usepackage{amssymb}
\usepackage{algorithmic}
\usepackage{algorithm}
\usepackage{array}
\usepackage[caption=false,font=normalsize,labelfont=sf,textfont=sf]{subfig}
\usepackage{float}
\usepackage{textcomp}
\usepackage{stfloats}
\usepackage{url,soul}
\usepackage{verbatim}
\usepackage{graphicx}
\usepackage{cite}
\newcommand{\bm}[1]{\mbox{\boldmath{$#1$}}}

\begin{document}

\title{Proximal Policy Optimization-based Transmit Beamforming and Phase-shift Design in an IRS-aided ISAC System for the THz Band}

	\author{Xiangnan Liu, Haijun Zhang,~\IEEEmembership{Senior Member,~IEEE}, Keping Long,~\IEEEmembership{Senior Member,~IEEE}, Mingyu Zhou, Yonghui Li,~\IEEEmembership{Fellow,~IEEE}, and H. Vincent Poor,~\IEEEmembership{Life Fellow,~IEEE}.
	\thanks{This work is supported in part by the National Key Research and Development Program of
		China (Grant No. 2020YFB1708800), in part by Beijing Natural Science Foundation (L212004),
		in part by the National Natural Science Foundation of China under Grants 61822104 and
		61771044, in part by the State Key Laboratory of Advanced Metallurgy under Grant KF20-04,
		in part by the Fundamental Research Funds for the Central Universities under Grants FRFTP-
		19-002C1 and RC1631, and in part by Beijing Top Discipline for Artificial Intelligent Science
		and Engineering, University of Science and Technology Beijing (\emph{Corresponding author: Haijun Zhang}.)
		
		Xiangnan Liu, Haijun Zhang, and Keping Long are with Institute of Artificial Intelligence, Beijing Advanced Innovation Center for Materials Genome Engineering, Beijing Engineering and Technology Research Center for Convergence Networks and Ubiquitous Services, University of Science and Technology Beijing, Beijing 100083, China (email: xiangnan.liu@xs.ustb.edu.cn, haijunzhang@ieee.org, longkeping@ustb.edu.cn).
		
		Mingyu Zhou is with BaiCells Technologies Company, Beijing, China, 100094 (email: Zhoumingyu@baicells.com).
		
		Yonghui Li is with the School of Electrical and Information Engineering, The University of Sydney, Sydney, NSW 2006, Australia (email:  yonghui.li@sydney.edu.au).
		
		H. Vincent Poor is with the Department of Electrical and Computer, Princeton University, NJ 08544 USA (e-mail: poor@princeton.edu).}}

\maketitle

\begin{abstract}
In this paper, an IRS-aided integrated sensing and communications (ISAC) system operating in the terahertz (THz) band is proposed to maximize the system capacity. Transmit beamforming and phase-shift design are transformed into a universal optimization problem with ergodic constraints. Then the joint optimization of transmit beamforming and phase-shift design is achieved by gradient-based, primal-dual proximal policy optimization (PPO) in the multi-user multiple-input single-output (MISO) scenario. Specifically, the actor part generates continuous transmit beamforming and the critic part takes charge of discrete phase shift design. Based on the MISO scenario, we investigate a distributed PPO (DPPO) framework with the concept of multi-threading learning in the multi-user multiple-input multiple-output (MIMO) scenario. Simulation results demonstrate the effectiveness of the primal-dual PPO algorithm and its multi-threading version in terms of transmit beamforming and phase-shift design.
\end{abstract}


	\section{Introduction}
\IEEEPARstart{R}{recently}, the integrated sensing and communications (ISAC) has been proposed to enhance the sensing capability in the location/environment-aware scenarios. Many scenarios require ISAC to obtain high-rate transmission and high-resolution target detection, such as  autonomous vehicles, indoor localization, and extended reality (XR) \cite{Reference1}. ISAC is considered to be a promising technique for the next generation wireless systems to support high-accuracy sensing services \cite{Reference2}. Following the communication signal processing, the millimeter wave (mmWave) band is considered in the ISAC system. Previous discussions of ISAC systems have considered the use of the mmWave bands for use in automotive radars and high-resolution imaging radars \cite{Reference3}. Therefore, higher bandwidth and transmission rate is a technique development in the envisioned  sixth generation (6G) networks.

In this context, terahertz (THz) communication is expected to enable ultra-high-speed communications in the era of 6G. 
The 6G networks will require higher data rates and capacity to maintain quality of service, economical operation, and flexible resource management \cite{Reference4}. The THz band can provide wider communication bandwidth than the current wireless communication band. THz communication can obtain Gbps wireless transmission rate, which will enable new sensing applications such as miniaturized radars for gesture detection and touchless smartphones, spectrometers for explosive detection and gas sensing.
THz communication can provide higher transmission rates and wider bandwidths than communication in lower-frequency bands. However, the molecular absorption in the THz band is severe. And the THz waves with strong directivity and poor diffraction are blocked by many obstacles more easily.

To tackle these challenges, the intelligent reflecting surface (IRS) has recently emerged as a promising technique to address the issue of excessive pathloss by creating better propagation environments \cite{Reference5}. IRSs are composed of passive antenna elements with adjustable phase shifts. Combined with a phase-shift design in an IRS, transmit beamforming can realize better ISAC performance. On the one hand, the transmit beamforming can be utilized to synthesize multiple beams towards existing users and targets \cite{Reference6}. 
On the other hand, the IRS can boost the ISAC signal by changing the phase optimization dynamically \cite{Reference7}. Compared with a conventional relay, the IRS does not employ a radio frequency unit and so can save energy. Thus an IRS-aided wireless communication system can obtain higher spectrum and energy efficiencies. 

By utilizing the IRS technique, the communication performance of the ISAC system can be enhanced. Meanwhile, high-data and high-resolution sensing performance can be achieved in the THz band. While IRS-aided systems open new possibilities, they also bring new design challenges, such as passive beamforming, channel realization, and deployment of the IRS. Moreover, there have been few studies of transmit beamforming and phase shift design for the IRS-aided ISAC system. Therefore. we focus on the joint optimization of transmit beamforming and phase-shift design in this paper.

\subsection{Related Works}
Transmit beamforming and phase-shift design in general IRS-aided system have attracted extensive attention in recent years. They have been studied in various systems, such as multi-user multiple-input single-output (MISO) \cite{Reference8}, multiple-input multiple-output (MIMO) communication \cite{Reference9}, and simultaneous wireless information and power transfer (SWIPT) systems \cite{Reference10}. To reduce the power consumption, Abeywickrama et.al. proposed a practical method by jointly designing the access point transmit beamforming and IRS passive phase optimization, subject to constraints on users’ individual signal-to-interference-plus-noise ratios (SINRs) \cite{Reference8}.
Transmit beamforming was studied in an IRS-assisted SWIPT system, with the constraints of quality-of-service (QoS) constraints at all users, by applying a penalty-based optimization method \cite{Reference10}. In \cite{Reference11}, Shen et al. first derived a closed-form solution for the base station (BS)'s transmit beamforming to maximize the signal-to-noise ratio (SNR) of a radar signal. Furthermore, transmit beamforming has been designed to optimize various objective functions, such as power consumption \cite{Reference12}, achievable rate \cite{Reference13}, and energy efficiency \cite{Reference14}. On the one hand, THz will enable new sensing applications such as miniaturized radars for gesture detection and touchless smartphones, spectrometers for explosive detection and gas sensing. On the other hand, IRS is a powerful relay with unique properties to enhance the radar and communication signal strength in the THz band. Therefore, the joint optimization of transmit beamforming and phase-shift design is meaningful, in terms of the communication and radar performance.

Deep reinforcement learning (DRL) is a state-of-the-art method that can be used to optimize radio resource allocation \cite{Reference15}. Despite the fact that excellent performances of many optimization algorithms have been observed through numerical simulations and theoretical analysis, implementing them in real systems still faces many serious difficulties. In particular, the high computational cost incurred by these algorithms has been one of the most challenging issues. Deep learning can reduce the computational cost but it acquires the training set, resulting in poor flexibility of deep learning. Furthermore, DRL-based methods do not need training labels and possess the property of online learning and sample generation, which is more storage-efficient. Recently, it has been leveraged to optimize performance in wireless systems extensively, for problems such as power control \cite{Reference16}, bandwidth scheduling \cite{Reference17}, computation offloading \cite{Reference18}, and caching deployment \cite{Reference19}. Mismar et al. proposed joint optimization of analog beamforming, power allocation, and interference coordination in the sub-6 GHz and mmWave band. A deep Q-learning network (DQN) was applied to tackle  this joint optimization \cite{Reference20}. Recently DRL has been applied to solve optimization problems in IRS assisted communications \cite{Reference21,Reference22,Reference23}. In \cite{Reference21}, a DRL framework is designed to investigate the passive phase optimization for the IRS-aided MISO scenario to maximize the SNR. The results demonstrated a considerable gain compared to existing schemes. In \cite{Reference22}, the transmit beamforming and passive phase were jointly optimized by using a DRL framework. The developed framework can support large-dimensional optimization problems in massive MIMO systems. The transmit beamforming in the BS and passive beamforming in the IRS were jointly designed to optimize the system transmission rate in \cite{Reference23} by using prioritized. Towards the complicated practical scenario, the DRL with a single thread is intractable because of its lower utilization of computing process.

Furthermore, distributed DRL is a promising technique to solve more complicated wireless networks problems. Distributed DRL is composed of a central controller and a group of learners. It can be classified as multi-agent DRL or multi-threading DRL. The latter uses the multi-threading method to train agents. Interactive learning takes place simultaneously in multiple threads and each thread summarizes the learning results, sorting and saving them in a public place. Compared with multi-agent DRL, multi-threading DRL can save power consumption. Additionally, multi-threading DRL can save training time and keep the training process stable \cite{Reference24}. Du et al. proposed an asynchronous advantage actor-critic (A3C) algorithm to solve joint optimization of viewport rendering offloading and power allocation \cite{Reference25}. Zhang et al. solved the optimization problem of power control in the cognitive radio network by utilizing distributed DRL methods, such as A3C and distributed proximal policy optimization (DPPO) \cite{Reference26}. Dinh et al. designed a distributed model-free algorithm DeepPool to optimize ride-sharing platforms \cite{Reference27}. Compared with multi-agent reinforcement learning, there are relatively few studies of multi-threaded DRL, especially for transmit beamforming and passive phase optimization. 

On the other hand, ergodic stochastic optimization, which uses term averages to reflect system performance,  is suitable for wireless resource allocation.
	This idea was originated by Ribeiro \cite{Reference28}, who proposed the average variable scheme for radio resource management. This ergodic system performance metric is valid for both independent channels and correlated channels.
	A further study \cite{Reference29} considered optimizing primal and dual variables by learning resource allocation policy gradients. Lee et al. in \cite{Reference30} proposed a distributed scheme involving average variables to capture system performance, and two deep neural networks (DNNs) were introduced to approximate the allocation policy. Combining this scheme of average variables with a DRL framework is a promising approach, which also becomes one of motivations of this paper.
\subsection{Contributions}

In this context, we study an IRS-aided ISAC system in the THz band. In particular, transmit beamforming and phase-shift design in an IRS-aided device-based ISAC system with multiple users are studied. The capacity with ergodic constraints is set as the utility function to evaluate the IRS-aided ISAC system performance. This problem is challenging because of the joint optimization of transmit beamforming and phase-shift design. To tackle this problem, proximal policy optimization (PPO) and its distributed forms are developed in different scenarios, including MISO and MIMO.
The main contributions of this paper are summarized below.
\begin{itemize}
	\item
	\emph{The IRS-aided ISAC system in the THz band:} The IRS is explored to compensate for the high path loss caused by molecular absorption in the THz band. We consider an IRS-aided ISAC system consisting one BS equipped with several antennas, several users and an IRS. The capacity of the ISAC system is optimized through transmit beamforming and passive phase-shift design of the IRS.
\end{itemize}

\begin{itemize}
	\item
	\emph{The primal-dual model-free PPO with ergodic constraints:} 
	The optimization problem of capacity maximization is formulated as a universal optimization problem with ergodic constraints and is transformed into the dual domain. A gradient-based, primal-dual PPO algorithm is designed to solve the problem of capacity maximization with ergodic constraints.
\end{itemize}

\begin{itemize}
	\item
	\emph{The joint optimization via the actor-critic structure:} 
	PPO is leveraged to achieve joint optimization on continuous transmit beamforming of the ISAC BS and discrete phase design of the IRS. The actor part optimizes transmit beamforming and the critic part designs the phase shifts, to maximize the total capacity in the multi-user MISO scenario.
\end{itemize}

\begin{itemize}
	\item
	\emph{The optimization of multi-user MIMO scenario through distributed DRL:} Distributed PPO (DPPO) is introduced in the established model to solve this joint optimization. The concept of multi-threading applied to the more complicated matrix processing in the MIMO scenario. Each worker docks with one user to collect observations and transmit them to the global PPO. The global PPO then broadcasts the optimized gradients to each worker.
\end{itemize}

The DRL-based method to achieve the joint optimization of transmit beamforming and phase-shift design is different from the current studies of transmit beamforming in ISAC system. Utilizing the IRS to counteract the high pathloss in the THz band, this emerging technique still lacks in-depth studies studies for ISAC system. Towards the joint optimization in the multi-user MISO scenario, the proposed primal-dual PPO algorithm is novel and effective. Beyond the multi-user MISO scenario, the primal-dual PPO algorithm's distributed version can be applied to the multi-user MIMO scenario. Simulation results will demonstrate the effectiveness of primal-dual PPO-based algorithm in both the multi-user MISO scenario and multi-user MIMO scenario.

The rest of this paper is organized as follows. Section II introduces the system model and formulates the optimization problem of transmit beamforming and phase-shift design. Section III presents the primal-dual PPO algorithm to realize joint optimization of transmit beamforming and passive phase  optimization in the multi-user MISO scenario. In Section IV, the primal-dual DPPO is extended to the MIMO scenario. Section V shows the pseudo codes and implementations of these two algorithms. The proposed algorithms are verified by simulations results in VI, and Section VII concludes the paper.
\section{System Model and Problem Formulation}
\subsection{System Model}
\begin{figure}[h]
	\centering
	\includegraphics[width=85mm]{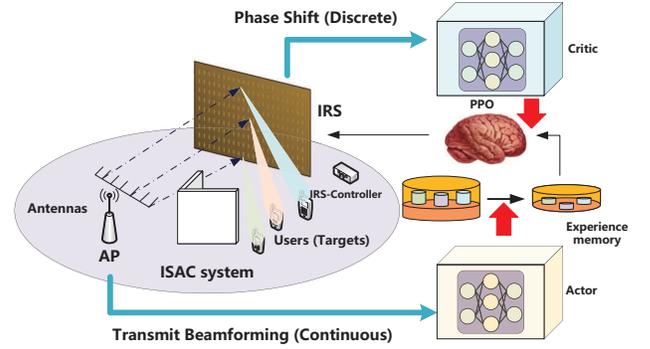}
	\caption{Primal-dual PPO learning beamforming and phase-shift design in the  IRS-aided ISAC system.}
	\label{fig1}
\end{figure}
As shown in Fig. 1, a multi-user MISO model is considered in the IRS-aided ISAC system.
It consists of one ISAC BS with $M$ antennas and $K$ users equipped with single antenna.
Let ${\cal M} = \left\{ {1,2,...,M} \right\}$ and ${\cal K} = \left\{ {1,2,...,K} \right\}$ denote the set of the BS's antennas and served users. ${\bf{s}} = {\left[ {{s_1},s_2^{},...,s_k^{}} \right]^T} \in {{\Bbb C}^{K \times 1}}$ and ${\bf{W}} \in {{\Bbb C}^{M \times K}}$ are the information-bearing symbol and the transmit beamforming matrix, respectively.
The transmit signal ${\bf{X}} \in {{\Bbb C}^{M \times 1}}$ after transmit beamforming is given by
\begin{equation}
		{\bf{X}} = {\bf{W}}\cdot{\bf{s}}.
\end{equation}

\emph{1) Communication Signal:}

Let ${{\bf{H}}_1} \in {{\Bbb C}^{N \times M}}$, ${{\bf{H}}_0} \in {\Bbb C}^{{M \times K}}$, and ${{\bf{H}}_2} \in {\Bbb C}^{{N \times K}}$ represent the channel gain from the ISAC BS to the IRS, and the channel gain from the ISAC BS to user $k$, the channel gain from the IRS's element $n$ to user $k$, respectively.
The communication signal ${{\bf{y}}_c} \in {{\Bbb C}^{K \times 1}}$ received by users can be written as
\begin{equation}
		{\bf{y}}_c = \left( {{{\bf{H}}_0^H} + {{\bf{H}}_2^H}{\bf{V}}{{\bf{H}}_1}} \right){{\bf{X}}^{}} + {\bf{n}}_c,
\end{equation}
${\bf{n}}_c^{} \in {{\Bbb C}^{K \times 1}}$ is the independent identically distributed complex Gaussian stochastic process. The effective diagonal phase matrix $\bf V$ is $diag\left\{ {{\gamma _1}{e^{j{\beta _1}}},{\gamma _2}{e^{j{\beta _2}}},...,{\gamma _N}{e^{j{\beta _N}}}} \right\}$, where ${\gamma _n} \in \left[ {0,1} \right]$ and ${\beta _n} \in \left[ {0,2\pi } \right]$ are amplitude and phase, respectively. Let ${\cal N} = \left\{ {1,2,...,N} \right\}$ denote the set of the IRS's elements.
Define $\gamma _n$ as the amplitude reflection coefficient of the IRS and it can be calculated as \cite{Reference8}
\begin{equation}
	{\gamma _n}\left( {{\beta _n}} \right) = \left( {1 - {\gamma _{\min }}} \right){\left( {\frac{{\sin \left( {{\beta _n} - \varphi } \right) + 1}}{2}} \right)^\varepsilon } + {\gamma _{\min }}.
\end{equation}
${\gamma _{\min }}$ is the minimum amplitude, $\varphi$ is the horizontal distance between $-\pi/2$ and ${\gamma _{\min }}$. $\varphi \ge 0$ and $\varepsilon  \ge 0$ are the constants depending on the IRS's circuit.

All of them are modeled as the Rician channel \cite{Reference31},

\begin{equation}
		{{\bf{H}}_{i}} = \sqrt {\frac{\gamma_{AI} }{{1 + \gamma_{AI}}}} {{\bf{H}}_{LOS}} + \sqrt {\frac{1}{{1 + \gamma_{AI} }}} {{\bf{H}}_{NLOS}},i=0,1,2,
\end{equation}
where $\gamma_{AI}$ is the Rician factor, ${\bf{H}}_{LOS}$ is the line-of-sight (LoS) component and it can be expressed as ${{\bf{H}}_{LOS}} = { {\alpha}_{LOS}\left( {f,l} \right){G_t}{G_r}{{\bf{b}}_r{\left( {{\theta_{AoA}}} \right)}}{\bf{a}}_t^{H}\left( {{\theta_{AoD}}} \right)}$,
where ${{\bf{b}}_r}\left( \theta_{AoA}  \right)$ and ${{\bf{a}}_t}\left( \theta_{AoD}  \right)$ 
is the receive response vector and transmit response vector, respectively. ${{\mathbf{a}}_t^{H}}\left( \cdot   \right)$ denotes the conjugate array steering vector.
Additionally, $\theta_{AoD}$ and $\theta_{AoA}$ are the angle of departure (AoD) and the angle of arrival (AoA), respectively.

We consider a uniform linear antenna (ULA) to be the structure of the ISAC BS's antennas, so the steering vector ${{\bf{a}}}\left( \theta  \right) \in {{\Bbb C}^{M \times 1}}$ is given by
\begin{equation}{ 
		{{\bf{a}}}\left( \theta  \right) = \sqrt {\frac{1}{M}} {\left[ {1,{e^{ - j{{2\pi d} \mathord{\left/
								{\vphantom {{2\pi d} \lambda }} \right.
								\kern-\nulldelimiterspace} \lambda }\cos \theta }},...,{e^{ - j{{2\pi d} \mathord{\left/
								{\vphantom {{2\pi d} \lambda }} \right.
								\kern-\nulldelimiterspace} \lambda }\left( {M - 1} \right)\cos \theta }}} \right]^T}.
	}
\end{equation}
$d$ is the antenna element spacing and $\lambda$ is the wavelength.

For the IRS, an $N \times N$ element uniform planar antenna (UPA) is considered. The array steering vector ${{\mathbf{v}}}\left( {\phi ,{\mathbf{\theta }}} \right) \in  {{\Bbb C}^{N \times 1}}$ is given by

\begin{equation}
		\begin{array}{*{20}{l}}
			{{{\mathbf{v}}}\left( {\phi ,{\mathbf{\theta }}} \right) = \frac{1}{N}\left[ {1,...,{e^{j\frac{{2\pi }}{\lambda }d\left( {n\cos {\phi}\cos {\theta} + n\sin {\phi}\sin {\theta}} \right)}},...{\kern 1pt} {\kern 1pt} ,} \right.} \\
			{{{\left. {{\kern 1pt} {\kern 1pt} {\kern 1pt} {\kern 1pt} {\kern 1pt} {\kern 1pt} {\kern 1pt} {\kern 1pt} {\kern 1pt} {\kern 1pt} {\kern 1pt} {\kern 1pt} {\kern 1pt} {\kern 1pt} {\kern 1pt} {\kern 1pt} {\kern 1pt} {\kern 1pt} {\kern 1pt} {\kern 1pt} {\kern 1pt} {\kern 1pt} {\kern 1pt} {\kern 1pt} {\kern 1pt} {\kern 1pt} {\kern 1pt} {\kern 1pt} {\kern 1pt} {\kern 1pt} {\kern 1pt} {\kern 1pt} {\kern 1pt} {\kern 1pt} {\kern 1pt} {\kern 1pt} {\kern 1pt} {\kern 1pt} {\kern 1pt} {\kern 1pt} {\kern 1pt} {e^{j\frac{{2\pi }}{\lambda }d\left( {\left( {\sqrt N  - 1} \right)\cos {\phi }\cos {\theta} + \left( {\sqrt N  - 1} \right)\sin {\phi}\sin {\theta}} \right)}}} \right]}^T}}.
	\end{array}
\end{equation}
$n$ is the antenna element index with $0 \le n \le N - 1$.

In the THz band, there is not only free space pathloss $L_{spread}$, but also the module pathloss $L_{medium}$. The pathloss of LoS ${\left| {{\alpha _L}\left( {f,l} \right)} \right|^2}$ is expressed by
\begin{equation}
	{\left| {{\alpha _L}\left( {f,l} \right)} \right|^2} = {L_{spread}}\left( {f,l} \right){L_{medium}}\left( {f,l} \right).
\end{equation}
$f$ is the carrier frequency and $l$ is the transmission distance. The free space pathoss $L_{spread}$ can be calculated by
\begin{equation}
	{L_{spread}}\left( {f,l} \right){\text{ = }}{\left( {\frac{c}{{4\pi fl}}} \right)^2}.
\end{equation}

Similar to the channel state of the LoS path ${{\bf{H}}_{LOS}}$, ${{\bf{H}}_{NLOS}} = {  \sum\limits_{i = 1}^{{n_{NL}}} {{\alpha _i}\left( {f,l} \right){G_t}{G_r}{{\bf{b}}_r}\left( {{\theta _{AoA,i}}} \right){\bf{a}}_t^{H}\left( {{\theta _{AoD,i}}} \right)}}$ is set as the non-line-of-sight (NLoS) component. Let $G_t$ and $G_r$ represent channel gain from transmit antenna and received antenna, respectively. ${n_{NL}}$ is the number of NLoS rays and $\alpha \left( {f,l} \right)$ is the complex gain of the path.

The large-scale channel gain $\alpha _{LOS}$ and $\alpha _{NLOS}$ can be given by
\begin{subequations}
	{\begin{align}
			&\begin{array}{*{20}{l}}{\alpha _{LOS}\left( {f,l} \right)} = {L_{spread}}\left( {f,l} \right){L_{abs}}\left( {f,l} \right)\\ {\kern 1pt} {\kern 1pt} {\kern 1pt} {\kern 1pt} {\kern 1pt} {\kern 1pt} {\kern 1pt} {\kern 1pt}{\kern 1pt} {\kern 1pt} {\kern 1pt} {\kern 1pt} {\kern 1pt} {\kern 1pt} {\kern 1pt} {\kern 1pt}{\kern 1pt} {\kern 1pt} {\kern 1pt} {\kern 1pt} {\kern 1pt} {\kern 1pt} {\kern 1pt} {\kern 1pt}{\kern 1pt} {\kern 1pt} {\kern 1pt} {\kern 1pt} {\kern 1pt} {\kern 1pt} {\kern 1pt} {\kern 1pt}{\kern 1pt} {\kern 1pt} {\kern 1pt} {\kern 1pt} {\kern 1pt} {\kern 1pt} {\kern 1pt} {\kern 1pt}{\kern 1pt} {\kern 1pt} {\kern 1pt} {\kern 1pt} {\kern 1pt} {\kern 1pt} {\kern 1pt} {\kern 1pt}{\kern 1pt} {\kern 1pt} {\kern 1pt} {\kern 1pt} {\kern 1pt}= {\left( {\frac{c}{{4\pi fl}}} \right)^2}{e^{ - {k_{abs}}\left( f \right)l}},\end{array}\\
			&\begin{array}{*{20}{l}}{\alpha _{NLOS}\left( {f,l} \right)} = {\Gamma ^2\left( f \right)}{L_{spread}}\left( {f,l} \right){L_{abs}}\left( {f,l} \right)\\ {\kern 1pt} {\kern 1pt} {\kern 1pt} {\kern 1pt} {\kern 1pt} {\kern 1pt} {\kern 1pt} {\kern 1pt}{\kern 1pt} {\kern 1pt} {\kern 1pt} {\kern 1pt} {\kern 1pt} {\kern 1pt} {\kern 1pt} {\kern 1pt}{\kern 1pt} {\kern 1pt} {\kern 1pt} {\kern 1pt} {\kern 1pt} {\kern 1pt} {\kern 1pt} {\kern 1pt}{\kern 1pt} {\kern 1pt} {\kern 1pt} {\kern 1pt} {\kern 1pt} {\kern 1pt} {\kern 1pt} {\kern 1pt}{\kern 1pt} {\kern 1pt} {\kern 1pt} {\kern 1pt} {\kern 1pt} {\kern 1pt} {\kern 1pt} {\kern 1pt}{\kern 1pt} {\kern 1pt} {\kern 1pt} {\kern 1pt} {\kern 1pt} {\kern 1pt} {\kern 1pt} {\kern 1pt}{\kern 1pt} {\kern 1pt} {\kern 1pt} {\kern 1pt} {\kern 1pt} = {\left( {\frac{c}{{4\pi fl}}} \right)^2}{e^{ - {k_{abs}}\left( f \right)l}},
			\end{array}
	\end{align}}
\end{subequations}
the speed of light is $c = 3 \times {10^8}\;{m \mathord{\left/ {\vphantom {m s}} \right. \kern-\nulldelimiterspace} s}$. Let $l$ denote the transmission distance in free space. The absorption coefficient $k_{abs}\left( f \right)$ can be calculated by the medium available at a molecular level \cite{Reference25}. The reflecting coefficient ${\Gamma}\left( f \right)$ is the product of the Fresnel reflection coefficient $\iota  $ and the Rayleigh roughness factor $\xi$.
The high reflection loss (i.e. up to second order reflections) is only considered in the THz band. The Fresnel reflection coefficient $\iota$ can describe the behavior of light in medium with different refractive indices. It can be calculated by
\begin{equation}
	\iota \left( f \right) = \frac{{Z\left( f \right)\cos {\varphi _{in}} - {Z_0}\cos {\varphi _{ref}}}}{{Z\left( f \right)\cos {\varphi _{in}} + {Z_0}\cos {\varphi _{ref}}}},
\end{equation}
where $Z\left(f\right)$ is the reflecting material 's frequency-dependent wave impedance and ${{\rm{Z}}_0}{\rm{  =  377}}{\kern 1pt} {\kern 1pt}{\kern 1pt} {\kern 1pt} \Omega {\rm{ }}$ is the free space wave impedance. Let $\varphi_{in}$ denote the angle of incidence and reflection. Additionally, ${\varphi _{ref}} = \arcsin \left( {{\textstyle{Z \over {{Z_0}}}}\sin {\varphi _{in}}} \right)$ represents the angle of refraction.

The Rayleigh roughness factor $\xi$ is calculated by
\begin{equation}
	\zeta \left( f \right) = {e^{ - {\textstyle{1 \over 2}}{{\left( {{\textstyle{{4\pi f\sigma \cos {\varphi _{in}}} \over c}}} \right)}^2}}}.
\end{equation}

They are composed of reflection and scattering, so each user $k$'s channel state can be calculated by \cite{Reference32}
\begin{equation}
		{\bf{y}}_{c,k} = \left( {{{\bf{H}}_{0,k}^H} + {{\bf{H}}_{2,k}^H}{\bf{V}}{{\bf{H}}_{1,k}}} \right){{\bf{X}}^{}} + {\bf{n}}_{c,k}={\bf H}_k{\bf X}_k+{\bf{n}}_{c,k},
\end{equation}
Furthermore, the user $k$ 's signal-to-interference-to-noise ratio (SINR)  can be calculated by
\begin{equation}
	{SINR}_k = \frac{{{{\left| {{\bf{H}}_k^H{{\bf{W}}_k}} \right|}^2}}}{{\sum\limits_{l \ne k} {{{\left| {{\bf{H}}_k^H{{\bf{W}}_l}} \right|}^2}}  + \sigma _c^2}}.
\end{equation}

\emph{2) Radar Signal:}

The radar signal received at the ISAC receiver is given by \cite{Reference33}
\begin{equation}
		{\bf{y}}_r^{} = \left( {{\bf{H}}_1^H{\bf{VB}}{{\bf{V}}^H}{{\bf{H}}_1}{\bf{ + A}}} \right){\bf{X}} + {\bf{n}}_r^{},
\end{equation}
It is assumed that IRS is a mono-static MIMO radar\cite{Reference2}. Thus, ${\bf{A}} \in {{\Bbb C}^{M \times M}}$ is the target response matrix of the ISAC BS and ${\bf{B}} \in {{\Bbb C}^{N \times N}}$ is the target response matrix based on parameters of the IRS's elements in the device-based ISAC system.
They can be calculated by,
\begin{subequations}
\begin{align}
			& {\bf{A}} = \sum\limits_{k = 1}^K {{\alpha _k}{{\bf{a}}_r}\left( {{\theta _k}} \right){{\bf{a}}_t^H}\left( {{\theta _k}} \right)},  \label{Za}\\
			& {\bf{B}} = \sum\limits_{k = 1}^K {{\alpha _k}{{\bf{v}}_r}\left( {{\upsilon_k ,\theta _k}} \right){\bf{v}}_t^H\left( {{\upsilon_k ,\theta _k}} \right)}. \label{Zb}
	\end{align}
\end{subequations}
$\alpha_k$ is the complex pathloss that include the coefficient of the pathloss, reflection, and complex radar cross of the target. Vector ${{\mathbf{a}}_t}\left( \theta  \right)$ and ${{\mathbf{a}}_r}\left( \theta  \right)$ are array steering vectors from transmit and receive antennas.
For radar signals, beam pattern error is considered in the proposed system \cite{Reference6}. MIMO radar transmit beamforming design aims to optimize the transmit power at given directions, or generally match a desired beam pattern.
The beam pattern error is the mean square error (MSE) between the obtained beam pattern and the ideal beam pattern. It can be calculated by the following formula
\begin{equation}
	{L_{r}}\left( {\bf{R}} \right) = \frac{1}{L}\sum\limits_{l = 1}^L {{{\left| {d\left( {{\psi _l}} \right) - P\left( {{\psi _l}} \right)} \right|}^2}},
\end{equation}
where ${\bf{R}} = {\bf{W}}{{\bf{W}}^H}$ is covariance matrix. The direction grids $\left\{ {{\psi _l}} \right\}_{l = 1}^L$ are sampled in the range of $ - 90^\circ $ to $  90^\circ $ with resolution of $ 0.1^\circ $. $P\left( \psi  \right) = {{\bf{a}}^H}\left( \psi  \right){\bf{Ra}}\left( \psi  \right)$ is power consumption in direction $\psi$ and $d\left( {{\psi}} \right)$ is desired beam pattern, its calculating principle is
\begin{equation}
	d\left( \psi \right) = \left\{ {\begin{array}{*{20}{l}}
			{1,}&{{\psi _p} - {\Delta  \mathord{\left/
						{\vphantom {\Delta  2}} \right.
						\kern-\nulldelimiterspace} 2} \le \psi  \le {\psi _p} + {\Delta  \mathord{\left/
						{\vphantom {\Delta  2}} \right.
						\kern-\nulldelimiterspace} 2},p = 1,2,3,}\\
			{0,}&{otherwise},
	\end{array}} \right.
\end{equation}
where the ideal beam pattern $\psi _p$ consists of three main beams and $\Delta = 10^\circ$ is each ideal beam width.

The beam pattern error is set as a constraint and it can be expressed as
\begin{equation}
	{L_{r}}\left( {\bf{R}} \right) \le \ell ,
\end{equation}
The above constraint implies that the beam pattern error is not allowed to exceed the threshold $\ell$. The threshold symbolizes the accuracy of the sensing performance.

\subsection{Problem Formulation}

The problem of transmit beamforming and phase shift design can be transformed into a long-term instantaneous performance function using the ergodic average value $\bf x$ to reflect the system performance \cite{Reference28}, and this solution can be applied to optimize other systems, including optimization of frequency division multiplexing, power control, and random access. It can be calculated by
\begin{equation}
	{\bf{x}} \le {\mathbb{E}}\left[ {{{\bf f}_1}\left( {{\bf{h}},{\bf{p}}\left( {\bf{h}} \right)} \right)} \right],
\end{equation}
where ${{\bf f}_1}\left( {{\bf{h}},{\bf{p}}\left( {\bf{h}} \right)} \right)$ is an instantaneous performance function. The design is to choose a resource allocation ${\bf{p}}\left( {\bf{h}} \right)$ to maximize the ergodic variable ${\bf{x}}$. The average variable ${\bf{x}}$ reflects the performance of wireless communication systems in a considerably long period and is influenced by instantaneous resource allocation ${\bf{p}}\left( {\bf{h}} \right)$.
\begin{equation}
		\begin{array}{*{20}{l}}
			{\max }&{{f_0}\left( {\mathbf{x}} \right)} \\ 
			{s.t.}&{{\bf x} \leqslant \mathbb{E}\left[ {{{\mathbf{f}}_1}\left( {{\mathbf{h}},{\mathbf{p}}\left( {\mathbf{h}} \right)} \right)} \right]} \\ 
			{}&{{{\mathbf{f}}_2}\left( {\mathbf{x}} \right) \geqslant 0,{\mathbf{x}} \in \chi ,{\mathbf{p}} \in \mathcal{P}.} 
	\end{array}
\end{equation}
In the proposed system model, ${\bf{h}}$ is the channel state $\bf H$, ${\bf{p}}\left( {\bf{h}} \right)$ is the transmit beamforming matrix $\bf W$ and phase optimization $\bf V$. ${{\bf{f}}_1}\left( {{\bf{h}},{\bf{p}}\left( {\bf{h}} \right)} \right)$ is the corresponding data transmission rate. ${{\bf{f}}_2}\left(\bf x\right)$ indicates the vector utility function.
The design goal is to maximize the average vector $\bf{x}$ for transmit  beamforming ${\bf{p}}\left( {\bf{h}} \right)$ under constraints,
\begin{equation}
	{x_k} \leqslant {{\Bbb E}_{\bf{H}}}\left[ {\log \left( {1 + SIN{R_k}} \right)} \right].
\end{equation}
Its design goal is to minimize radar signal loss performance with long-term constraints for transmit beamforming and phase-shift design. It is assumed that the element of covariance matrix ${\left[\bf{R}\right]_{mm}}$ has power budget:
\begin{equation}
		{{\left[\bf{R}\right]_{mm}}} \leqslant {{{P_{max}}} \mathord{\left/
				{\vphantom {{{P_{max}}} M}} \right.
				\kern-\nulldelimiterspace} M},m = 1,2,...,M,
\end{equation}
where $P_{max}$ is the total transmit power.
	The utility ${f_0}\left( {\bf{x}} \right)$ is set as the sum rate function.
\begin{equation}
	{f_0}\left( {\bf{x}} \right) = \sum\limits_{k = 1}^K {{x_k}}. 
\end{equation}

Finally, the optimization problem of MISO scenario in the IRS-aided ISAC system can be formulated as
\begin{equation}
		\begin{array}{*{20}{l}}
			{\mathop {\max }\limits_{{\bf{V,W}}} {f_0}\left( {\bf{x}} \right)}\\
			{\begin{array}{*{20}{l}}
					{s.t.{\kern 1pt} {\kern 1pt} {\kern 1pt} {\kern 1pt} {\kern 1pt} {\kern 1pt} {\kern 1pt} {x_k} \leqslant {{\Bbb E}_{\bf{H}}}\left[ {\log \left( {1 + SIN{R_k}} \right)} \right]},\\
					{{\kern 1pt} {\kern 1pt} {\kern 1pt} {\kern 1pt} {\kern 1pt} {\kern 1pt} {\kern 1pt} {\kern 1pt} {\kern 1pt} {\kern 1pt} {\kern 1pt} {\kern 1pt} {\kern 1pt} {\kern 1pt} {\kern 1pt} {\kern 1pt} {\kern 1pt} {\kern 1pt} {\kern 1pt} {\kern 1pt} {{\left[\bf{R}\right]_{mm}}} \le {{{P_{max}}} \mathord{\left/
								{\vphantom {{{P_{max}}} M}} \right.
								\kern-\nulldelimiterspace} M},m = 1,2,...,M,}\\
					{{\kern 1pt} {\kern 1pt} {\kern 1pt} {\kern 1pt} {\kern 1pt} {\kern 1pt} {\kern 1pt} {\kern 1pt} {\kern 1pt} {\kern 1pt} {\kern 1pt} {\kern 1pt} {\kern 1pt} {\kern 1pt} {\kern 1pt} {\kern 1pt} {\kern 1pt} {\kern 1pt} {\kern 1pt} {\kern 1pt} {L_{r}}\left( {\bf{R}} \right)\le \ell  }.
			\end{array}}
	\end{array}
\end{equation}

\section{Primal-dual Proximal Policy Optimization Scheme in Multi-user MISO Scenario}
In this section, we investigate multi-user MISO scenario in the IRS-aided ISAC system in the THz band, and use the primal-dual optimization to solve the problem (24).
\subsection{Transmit Beamforming and Phase-shift Design via PPO}
For the problem (24), the value-based method in DRL is not able to deal with continuous actions. The value-based method is unable to obtain the optimal solution with the constraints. Therefore, the research is mainly based on policy optimization.
The traditional policy-based solution is expressed by \cite{Reference29}:
\begin{equation}
	{L^{PG}}\left( {\bm{\omega }} \right) = {\mathbb{E}_t}\left[ {\log {\bm{\pi }}\left( {{\bf{H},\bm{\omega }}} \right){\mathbf{A}}\left( {{\bf{H},\bm{\omega }}} \right)} \right].
\end{equation}
The advantage function can be calculated by ${\bf{A}}\left( {{\bf{H},\bm{\omega }}} \right) = \sum\nolimits_{t' > t}^{} {{\gamma ^{t' - t}}{r^{t'}}}  - {v_\upsilon }\left( {{{\bf{H}}^t},{\bm \upsilon}^t} \right)$.
The shortcoming of this conventional method is the need to update the step size. Inappropriate setting of the step size will lead to a degraded performance. 
We need find a step size to ensure that the reward function increases monotonically for each iteration.	
On the other hand, the conventional policy gradient methods are sensitive to their hyper parameters and have high variance. Employing a trust region constraint can be regarded as an effective choice. The popular solution is trust region policy optimization (TRPO). It can be given by

\begin{equation}
	\begin{array}{*{20}{l}}
		{\mathop {\max }\limits_{\bm{\omega }} }&{{\mathbb{E}_{{\rho _\omega }\left( \tau  \right)}}\left[ {\sum\nolimits_t {{\gamma ^{t - 1}}\frac{{{\bm{\pi }}\left( {{\mathbf{H}},{\bm{\omega }}} \right)}}{{{{\bm{\pi }}_{old}}\left( {{\mathbf{H}},{\bm{\omega }}} \right)}}{{\mathbf{A}}_{old}}\left( {{\mathbf{H}},{\bm{\omega }}} \right)} } \right]}, \\ 
		{s.t.}&{{D_{KL}}\left[ {\frac{{{\bm{\pi }}\left( {{\mathbf{H}},{\bm{\omega }}} \right)}}{{{{\bm{\pi }}_{old}}\left( {{\mathbf{H}},{\bm{\omega }}} \right)}}} \right] \leqslant \delta } .
	\end{array}
\end{equation}

More detailed classifications on TRPO are presented in Appendix A.
The PPO algorithm is an approximate version of TRPO based on the first order gradients, utilizing DNNs to a large-scale distributed setting. PPO solves this problem by introducing relative entropy in policy update \cite{Reference34}.
\begin{equation}
	\begin{array}{*{20}{l}}
		{{L^{PPO}}\left( {\bm{\omega }} \right) = \sum\nolimits_t {\left( {{{{\bm{\pi }}\left( {{\bf{H},\bm{\omega }}} \right)} \mathord{\left/
							{\vphantom {{{\mathbf{\pi }}\left( {{\mathbf{H,\omega }}} \right)} {{{\mathbf{\pi }}_{old}}\left( {{\mathbf{H,\omega }}} \right)}}} \right.
							\kern-\nulldelimiterspace} {{{\bm{\pi }}_{old}}\left( {{\bf{H},\bm{\omega }}} \right)}}} \right){\mathbf{A}}\left( {{\bf{H},\bm{\omega }}} \right)} } \\ 
		{{\kern 1pt} {\kern 1pt} {\kern 1pt} {\kern 1pt} {\kern 1pt} {\kern 1pt} {\kern 1pt} {\kern 1pt} {\kern 1pt} {\kern 1pt} {\kern 1pt} {\kern 1pt} {\kern 1pt} {\kern 1pt} {\kern 1pt} {\kern 1pt} {\kern 1pt} {\kern 1pt} {\kern 1pt} {\kern 1pt} {\kern 1pt} {\kern 1pt} {\kern 1pt} {\kern 1pt} {\kern 1pt} {\kern 1pt} {\kern 1pt} {\kern 1pt} {\kern 1pt} {\kern 1pt} {\kern 1pt} {\kern 1pt} {\kern 1pt} {\kern 1pt} {\kern 1pt} {\kern 1pt} {\kern 1pt} {\kern 1pt} {\kern 1pt} {\kern 1pt} {\kern 1pt} {\kern 1pt} {\kern 1pt} {\kern 1pt} {\kern 1pt} {\kern 1pt} {\kern 1pt} {\kern 1pt} {\kern 1pt} {\kern 1pt}  - \lambda {D_{KL}}\left[ {{{{\bm{\pi }}\left( {{\bf{H},\bm{\omega }}} \right)} \mathord{\left/
						{\vphantom {{{\bm{\pi }}\left( {{\bf{H},\bm{\omega }}} \right)} {{{\bm{\pi }}_{old}}\left( {{\bf{H},\bm{\omega }}} \right)}}} \right.
						\kern-\nulldelimiterspace} {{{\bm{\pi }}_{old}}\left( {{\bf{H},\bm{\omega }}} \right)}}} \right]}.
	\end{array}
\end{equation}
Generally speaking, PPO is based on the actor-critic structure, and the actor part's goal is to maximize ${L^{PPO}}\left( \bm{\omega}  \right)$, which can be divided into two types, the one is ${L^{KLPEN}}\left(\bm \omega  \right)$ in PPO$_1$ and the other is ${L^{CLIP}}\left( \bm{\omega}  \right)$ in PPO$_2$. A penalty on Kullback-Leibler divergence (KL divergence) is used in PPO$_1$, and to adapt the penalty coefficient so that we achieve some target values of the KL divergence target each policy update. PPO$_2$ has no KL divergence term in the target, nor any constraints. Instead, it relies on tailoring the objective function to eliminate the incentives of the new policy and the old policy.
\begin{equation}
\begin{array}{*{20}{l}}
	{{L^{CLIP}}\left( {\bm \omega}  \right) = }&{{\mathbb{E}_t}\left[ {\min \left( {{\textstyle{{\pi \left( {H,{\bm \omega} } \right)} \over {{\pi _{old}}\left( {\bf H,{\bm \omega} } \right)}}}{\bf A}\left( {{\bf H},{\bm \omega} } \right),} \right.} \right.}\\
	{}&{\left. clip\left( {{\textstyle{{\pi \left( {{\bf H},{\bm \omega} } \right)} \over {{\pi _{old}}\left( {{\bf H},\omega } \right)}}},1 - \epsilon, 1 + \epsilon } \right){\bf A}\left( {{\bf H},{\bm \omega} } \right)\right]}.
\end{array}
\end{equation}
To get the similarity degree of the probability distribution of actions, KL divergence $D_{KL}$ can be used to calculate. Under the setting of hyperparameter is $\epsilon =2$, the second term $clip\left( {\frac{{\pi \left( {{\bf{H},\bm{\omega }}} \right)}}{{{\pi _{old}}\left( {{\bf{H,\bm{\omega} }}} \right)}},1 - \epsilon ,1 + \epsilon } \right)$ modifies the surrogate objective by clipping the probability ratio. 
In the formula (28), $\frac{{\pi \left( {{\bf{H},\bm{\omega }}} \right)}}{{{\pi _{old}}\left( {{\bf{H,\bm{\omega} }}} \right)}}$ can ensure that the distribution gap between the two updates is small. This clipped operation can avoid unnecessary samples to save training time.

To obtain optimization of transmit beamforming and phase-shift design, the actor part takes charge of continuous transmit beamforming in the established PPO framework. Meanwhile, the critic part undertakes the discrete phase-shift design.

The policy ${\bm{\pi }}\left( {{\bf{H},\bm{\omega }}} \right)$ is introduced to act as the transmit beamforming $\bf W$. For trained parameters $\bm{\omega }$, we make ${\bm{\pi }}\left( {{\bf{H},\bm{\omega }}} \right)={\bf W}$. And the problem (20) is transformed into the following problem.

\begin{equation}
		\begin{array}{*{20}{l}}
			{\mathop {\max }\limits_{{{\mathbf{H}}},{{\bm{\omega }}}} {f_0}\left( {{{\mathbf{x}}}} \right)} \\ 
			{\begin{array}{*{20}{l}}
					{s.t.{\kern 1pt} {\kern 1pt} {\kern 1pt} {\kern 1pt} {\kern 1pt} {\kern 1pt} {\kern 1pt} {{\mathbf{x}}} \leqslant \mathbb{E}\left[ {{{{\bf f}_1}}\left( {{\bf{H}},{\bm{\pi }}\left( {{\bf{H}},{\bm{\omega }}} \right)} \right)} \right]}, \\ 
					{{\mathbf{f}}_2}\left( {\mathbf{x}} \right) \geqslant 0. 
					
			\end{array}} 
	\end{array}
\end{equation}
The problem (29) is the general form of the problem (24). The constraint ${{\left[\bf{R}\right]_{mm}}} \le {{{P_{max}}} \mathord{\left/
			{\vphantom {{{P_{max}}} M}} \right.
			\kern-\nulldelimiterspace} M}$ is clipped by the action output of policy ${\bm{\pi }}\left( {{\bf{H},\bm{\omega }}} \right)$. We assume that ${{\mathbf{f}}_2}\left( {\mathbf{x}} \right)=\ell-{L_{r}}\left( {\bf{R}} \right)$. To convenience of understanding, the following formulation is based on the general form (29).

For the critic part, the phase shift of the instantaneous performance function can be calculated by the output action. The critic part adopts the design concept of DQN, it needs to be discrete, ${i_n} \in \left\{ {0,1,...,{2^b} - 1} \right\}$, the corresponding phase degree ${\beta _n}$ is given by
\begin{equation}
	{\beta _n} = \frac{{{i_n}2\pi }}{{{2^b} - 1}}.
\end{equation}
Accordingly, its Q-value function produces a discrete action ${i_n}$ as
\begin{equation}
	{i_n} = \mathop {\arg \max }\limits_{{{\bf{i}}_n}} {\bf Q}^{t}\left( { {{{\bf H}_k^{t}}},{{\bf{i}}_n}},{{\bm \upsilon}_k^t} \right).
\end{equation}
And the amplitude $\gamma_n$ also can be calculated through the formula (3). Therefore, the phase shift design of IRS can be obtained by the critic part.
\subsection{Primal-dual Learning Optimization}
The primal-dual optimization is introduced to tackle the problem (29). The Lagrangian of the problem (29) is utility and constraints weighted by their multipliers. For convenience, its Lagrangian function $\mathcal{L}\left(\cdot\right)$ for the problem (29) is given by
\begin{equation}
	\begin{array}{*{20}{l}}
		{\mathcal{L}\left( {{\bm{\omega }},{\bf{x}},{\bm{\lambda }},{\bm{\mu }}} \right) = {f_0}\left( {\bf{x}} \right) + {{\bm{\mu }}^T}{{{\bf f}_2}}\left( {\bf{x}} \right)}\\
		{{\kern 1pt} {\kern 1pt} {\kern 1pt} {\kern 1pt} {\kern 1pt} {\kern 1pt} {\kern 1pt} {\kern 1pt} {\kern 1pt} {\kern 1pt} {\kern 1pt} {\kern 1pt} {\kern 1pt} {\kern 1pt} {\kern 1pt} {\kern 1pt} {\kern 1pt} {\kern 1pt} {\kern 1pt} {\kern 1pt} {\kern 1pt} {\kern 1pt} {\kern 1pt} {\kern 1pt} {\kern 1pt} {\kern 1pt} {\kern 1pt} {\kern 1pt} {\kern 1pt} {\kern 1pt} {\kern 1pt} {\kern 1pt} {\kern 1pt} {\kern 1pt} {\kern 1pt} {\kern 1pt} {\kern 1pt} {\kern 1pt} {\kern 1pt} {\kern 1pt} {\kern 1pt} {\kern 1pt} {\kern 1pt} {\kern 1pt} {\kern 1pt} {\kern 1pt} {\kern 1pt} {\kern 1pt} {\kern 1pt} {\kern 1pt} {\kern 1pt} {\kern 1pt} {\kern 1pt} {\kern 1pt} {\kern 1pt} {\kern 1pt} {\kern 1pt} {\kern 1pt} {\kern 1pt} {\kern 1pt} {\kern 1pt} {\kern 1pt} {\kern 1pt} + {{\bm{\lambda }}^T}\left( {{\mathbb{E}}\left[ {{{{\bf f}_1}}\left( {{\bf{H}},{\bm{\pi }}\left( {{\bf{H}},{\bm{\omega }}} \right)} \right)} \right] - {\bf{x}}} \right)}.
	\end{array}
\end{equation}
Find the gradient of the four parameters in sequence,
\begin{equation}
	{{\bm{\omega }}^{t + 1}} = {{\bm{\omega }}^t} + {\tau _1}{\nabla _\omega }{\mathbb{E}}\left[ {{{\bf{f}}_1}\left( {{\bf{H}},{\bm{\pi }}\left( {{\bf{H}},{\bm{\omega }}} \right)} \right){{\bm{\lambda }}^t}} \right],
\end{equation}
\begin{equation}
	{{\bf{x}}^{t + 1}} = {{\bf{x}}^t} + {\tau _2}\left( {\nabla {f_0}\left( {{{\bf{x}}^t}} \right) + \nabla {{\bf{f}}_2}\left( {{{\bf{x}}^t}} \right){{\bm{\mu }}^t} - 1} \right),
\end{equation}
where $\tau _1, \tau _2 > 0$ denote scalar step sizes. A gradient update is performed on current dual iterates $\lambda _k, \mu _k$ in a similar manner by subtracting the partial stochastic gradients $\nabla _{\lambda} L$, $\nabla _{\mu} L$ and projecting onto the positive orthant to obtain
\begin{equation}
	{{\bm{\lambda }}^{t + 1}} = {{\bm{\lambda }}^t} - {\tau _3}\left( {{\mathbb{E}_{\bf H}}{{\bf{f}}_1}\left( {\bf{H}}, {{\bm{\pi }}\left( {{\bf{H}},{{\bm{\omega }}^{t + 1}}} \right)} \right) - {{\bf{x}}^{t + 1}}} \right).
\end{equation}
\begin{equation}
	{{\bm{\mu }}^{t + 1}} = {{\bm{\mu }}^t} - {\tau _4}{{\bf{f}}_2}\left( {{{\bf x}^{t + 1}}} \right).
\end{equation}
with associated step sizes $\tau _3, \tau _4 > 0$. The gradient primal-dual updates in (33)--(36) successively optimize the primal and dual variables to the maximum and minimum points of the Lagrangian function, respectively.

The updates in the formula (33) and (34) can be replaced with the zeroth-ordered gradient updates.
The gradients estimates with finite difference can be established using function observations in the given initial variables ${{\bm x}_0, {\bm \omega}_0}$ and the sampled points ${{\bf \widehat x}_1, {\bf \widehat x}_2, {\bm {\widehat \omega}}}$ as,
\begin{equation}
	\widehat \nabla {{{f}}_0}\left( {{{\bf{x}}_0}} \right): = \frac{{{{\widehat f}_0}\left( {{{\bf{x}}_0} + {\alpha _1}{{{\bf{\hat x}}}_1}} \right) - {{\widehat f}_0}\left( {{{\bf{x}}_0}} \right)}}{{{\alpha _1}}}{{\bf{\hat x}}_1},
\end{equation}
\begin{equation}
	\widehat \nabla {\bf{f}_2}\left( {{{\bf{x}}_0}} \right): = \frac{{{{\widehat {\bf{f}}}_2}\left( {{{\bf{x}}_0} + {\alpha _2}{{{\bf{\hat x}}}_2}} \right) - {{\widehat {\bf{f}}}_2}\left( {{{\bf{x}}_0}} \right)}}{{{\alpha _2}}}\widehat {\bf{x}}_2^T,
\end{equation}
\begin{equation}
	\begin{array}{*{20}{l}}
		{{{\widehat \nabla }_{\bm{\omega }}}\mathbb{E}\left[ {{\bf{f}_1}\left( {{\bm{\pi }}\left( {{\bf{H}},{\bm{\omega }}} \right),{\bf{H}}} \right)} \right]}
		\\{\kern 1pt} {\kern 1pt} {\kern 1pt} {\kern 1pt} {\kern 1pt} {\kern 1pt} {\kern 1pt} {\kern 1pt} {\kern 1pt} {\kern 1pt} {\kern 1pt} {\kern 1pt} {\kern 1pt} {\kern 1pt} {\kern 1pt} {\kern 1pt} {\kern 1pt} {\kern 1pt} {\kern 1pt} {\kern 1pt} {\kern 1pt} {\kern 1pt} {\kern 1pt} {\kern 1pt} {\kern 1pt} {\kern 1pt} {\kern 1pt} {\kern 1pt} {\kern 1pt} {\kern 1pt} {\kern 1pt} {\kern 1pt} {\kern 1pt} {\kern 1pt} {\kern 1pt} {\kern 1pt} {\kern 1pt} {\kern 1pt} {\kern 1pt} {\kern 1pt} {\kern 1pt} {\kern 1pt} {\kern 1pt} {\kern 1pt} {\kern 1pt} {\kern 1pt}{: = \frac{{\widehat {\bf{f}_1}\left( {{\bf{H}},{\bm{\pi }}\left( { {\bf{H}},{{\bm \omega}_0} + {\alpha _3}\widehat {\bm{\omega }}} \right)} \right) - \widehat {{\bf f}_1}\left( { {\bf{H}},{\bm{\pi }}\left( { {\bf{H}},{{\bm \omega_0 }}} \right)} \right)}}{{{\alpha _3}}}{{\widehat {\bm{\omega }}}^T}},
	\end{array}
\end{equation}
where $\alpha_1$, $\alpha_2$, $\alpha_3$ are the gradient estimated upgrading step factors. Furthermore, the $\tau_1$, $\tau_2$, $\tau_3$, $\tau_4$ are the primal-dual upgrading steps.

\section{Distributed Proximal Policy Optimization Scheme for Multi-user MIMO Scenario}
In this section, we will explore how to obtain joint optimization of transmit beamforming and phase shift design in the multi-user MIMO scenario. Distributed DRL will be introduced in this section.
\subsection{System Model and Formulation in Multi-user MIMO Scenario}
In the multi-user MIMO ISAC system, it is assumed that there exists a BS with $M$ transmit antennas and $K$ users with $R$ receive antennas. The transmit signal to the user $k$ through beamforming ${{\mathbf{X}}_k} \in {\mathbb{C}^{M \times R}}$ is  
\begin{equation}
		{{\mathbf{X}}_k} = {{\mathbf{W}}_k} \cdot {{\mathbf{s}}_k},
\end{equation}
where ${{\mathbf{s}}_k} = {\left[ {{s_1},{s_2},...,{s_R}} \right]^T} \in {\mathbb{C}^{R \times 1}}$. The responding transmit beamforming matrix is ${{\mathbf{W}}_k} \in {\mathbb{C}^{M \times R}}$ and radar signal is expressed by:
\begin{equation}
		{{\mathbf{y}}_{r,k}} = \left( {\mathbf{H}}_{1}^H{{\mathbf{V}}}{\mathbf{B}}{{\mathbf{V}}^H}{{\mathbf{H}}_{1}+{\mathbf{A}_k}} \right){{\mathbf{X}}_k} + {\mathbf{n}}_{r,k}.
\end{equation}
Similar to the formula (2), the communication signal ${\mathbf{y}}_{c,k}^{}$ is derived from
\begin{equation}
		{\mathbf{y}}_{c,k}^{} = \left( {{{\mathbf{H}}_{0,k}} + {{\mathbf{H}}_{2,k}}{\mathbf{V}}{{\mathbf{H}}_{1}}} \right){{\mathbf{X}}_k} + {\mathbf{n}}_{c,k}^{}.
\end{equation}
Compared with ${{\bf{H}}_0} \in {\Bbb C}^{{M \times K}}$ and ${{\bf{H}}_2} \in {\Bbb C}^{{N \times K}}$ in multi-user MISO scenario, the channel gains are transformed into ${{\mathbf{H}}_{0,k}} \in {\mathbb{C}^{M \times R}}$ and ${{\mathbf{H}}_{2,k}} \in {\mathbb{C}^{N \times R}}$, respectively.
And the corresponding parameterization problem is expressed as follows:
\begin{equation}
		\begin{array}{*{20}{l}}
			{\mathop {\max }\limits_{{{\mathbf{H}}_k},{{\mathbf{\omega }}_k}} {f_0}\left( {{{\mathbf{x}}_k}} \right)} \\ 
			{\begin{array}{*{20}{l}}
					{s.t.{\kern 1pt} {\kern 1pt} {\kern 1pt} {\kern 1pt} {\kern 1pt} {\kern 1pt} {\kern 1pt} {{\mathbf{x}}_k} \leqslant \mathbb{E}\left[ {{\text{log}}\left( {1 + SIN{R_k}\left( {{{\mathbf{H}}_k},{{{\bm \pi }}_k}\left( {{{\mathbf{H}}_k},{{{\bm \omega }}_k}} \right)} \right)} \right)} \right]}, \\ 
					{{\kern 1pt} {\kern 1pt} {\kern 1pt} {\kern 1pt} {\kern 1pt} {\kern 1pt} {\kern 1pt} {\kern 1pt} {\kern 1pt} {\kern 1pt} {\kern 1pt} {\kern 1pt} {\kern 1pt} {\kern 1pt} {\kern 1pt} {\kern 1pt} {\kern 1pt} {\kern 1pt} {\kern 1pt} {\kern 1pt} {{\left[{\bf R}_k\right]_{mm}}} \leqslant {{{P_{max}}} \mathord{\left/
								{\vphantom {{{P_{max}}} M}} \right.
								\kern-\nulldelimiterspace} M},m = 1,2,...,M}, \\ 
					{{\kern 1pt} {\kern 1pt} {\kern 1pt} {\kern 1pt} {\kern 1pt} {\kern 1pt} {\kern 1pt} {\kern 1pt} {\kern 1pt} {\kern 1pt} {\kern 1pt} {\kern 1pt} {\kern 1pt} {\kern 1pt} {\kern 1pt} {\kern 1pt} {\kern 1pt} {\kern 1pt} {\kern 1pt} {\kern 1pt} {L_{r}}\left( {{\mathbf R}_k} \right) \leqslant \ell . } 
			\end{array}} 
	\end{array}
\end{equation}
Similar to the problem (29), ${f_0}\left( {{{\mathbf{x}}_k}} \right)$ in the problem (43) is set as the sum rate function.
	Typically, the instantaneous system performance in practical MIMO transmission designs can not reflect the system performance well. One of solutions adopts the long term average such as ergodic capacity for transmission optimization \cite{Reference35}.

The channel gain ${\bf H}_k$ is the state in the environment, what we optimize is the transmit beamforming and phase-shift design. In this paper, the transmit beamforming ${\bf W}_k$ and phase-shift design ${\bf V}$ is policy derived form the state, i.e. channel gain ${\bf H}_k$. Similar to the multi-user MISO scenario, the ${{\bm \pi}_k\left({\bf H}_k,{\bm \omega}_k\right)}$ form actor part realizes the transmit beamforming and the critic part takes charge of the phase-shift design. When the scenario extended to the MIMO one, the number of receive antennas $R$ takes place of the number of users $K$. The channel ${\bf H}\in {\Bbb C}^{{M \times R \times K}}$ is 3 dimensions, compared with 2-dimension ${\bf H}$ in MISO scenario. How to solve higher dimension in multi-user MIMO IRS-aided ISAC system is worthy of to be studied, so DPPO, this multi-threading DRL technique is introduced in the following subsection.

\subsection{DPPO Optimization via multi-threading DRL}
DPPO with multiple threads can be adapted to more complex environments. DPPO is not a specific algorithm like PG or DQN, it conveys the concept of multi-threading learning. Its core is that in the training process of DRL, we can train multiple agents in parallel threading. In the training process, each worker takes charge of data collection locally.
Multiple workers can enrich the experiences form the environment. After the same episodes, distributed learning can accelerate the learning speed further because of its richer experience replay and its higher utilization rate of CPU.

The DPPO is suitable for the proposed multi-user MIMO scenario in IRS-aided ISAC system. Every user can act as each worker in multi-threading DRL. It can collect different channel realization from its own situation. Accordingly, each worker (i.e. user)  takes the corresponding policy to optimize its capacity. 

Through the total rewards from different workers (i.e. users), the original PPO algorithm can be achieved by estimated advantage functions ${{\bf A}}\left( {{\bf{H},\bm{\omega }}} \right)$. To facilitate the use of DNNs with batch updates while also supporting variable length episodes, L-step returns is proposed to estimate the advantage, i.e. we sum the rewards over the same L-step windows and bootstrap from the value function after L-steps: 
\begin{small}
\begin{equation}
	{{\bf A}^t}\left( {{\bf{H},\bm{\omega }}} \right) = \sum\nolimits_{l = 1}^L {{\gamma ^{l - 1}}{r^t} + } {\gamma ^{L - 1}}{v}\left( {{{\bf{H}}^{t + L}}},{{\bm \upsilon}^{t + L}} \right) - {v }\left( {{\bf{H}}^t},{{\bm \upsilon}^t} \right).
\end{equation}
\end{small}
Workers share a global PPO and the gradients of PPO would not be calculated by workers. Gradients will not be transformed to the chief like A3C. Workers only push the data collection by themselves to the global PPO. Each worker calculates the local $L^{CLIP}_k\left(\bm{\omega}\right)$ by the formula (29), computes gradient ${\nabla _\omega }{L^{CLIP}}$ and sends to the global PPO. The global PPO updates a certain batch of data from multiple workers (workers stop collecting when the chief is updating). After updating, workers collect data with the latest policy ${\bm \pi}^t$.

\subsection{Transmit Beamforming and Phase-shift Design in Multi-user MIMO Scenario}
\begin{figure}[h]
	\centering
	\includegraphics[width=85mm]{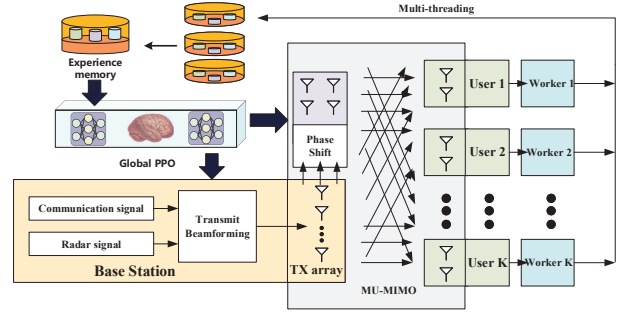}
	\caption{Primal-dual DPPO beamforming and phase-shift design in the  multi-user MIMO scenario.}
	\label{fig1}
\end{figure}
In the design of the proposed multi-user MIMO scenario, the dimension of channel state between the ISAC BS and users consists of $M \times R \times K$. It is quite difficult to realize transmit beamforming and passive phase optimization directly. With the help of DPPO's multi-threading training, each user will be allocated to each worker. The workers can collect observation and choose actions in parallel. Compared with multi-user MISO scenario, each worker processes the channel state ${\bf H}_k \in  {\Bbb C}^{{M \times R}}$, which is similar to ${\bf H} \in  {\Bbb C}^{{M \times K}}$. The global PPO will allocate actions to different workers according to the experience memory.

\section{Transmit Beamforming and Phase-Shift Design Optimization Algorithm Design}
\subsection{Primal-dual PPO Algorithm for Multi-user MISO Scenario}
In the proposed algorithm, we utilize the PPO's architecture to learn two kinds of actions. Firstly, we allocate the equal initial beamforming to each user, and initial policy parameters including mean $\bm{\omega}^{0}$ and Langrange multipliers $\bm{\lambda}^{0}$ and $\bm{\mu}^{0}$.
And then, each of the ISAC BS's antennas deploys PPO method in turn, and the sum of antennas' iteration is $M$. Next, beginning with the actor part's and the critic part's training, we compute the gradient estimate from replay experience. The estimated gradient integrates into the primal-dual optimization. The primal-dual PPO algorithm is used to iteratively calculate the  transmit beamforming matrix ${\bf W}^{\rm *}$ and the phase-shift design matrix ${\bf V}^{\rm *}$.

In Algorithm 1, episode begins with channel state ${\bf H}^0 \in {\Bbb C}^{{M \times K}}$ between the ISAC BS and users. It can be calculated by the formula (4) and the first phase shift ${\bf V}^0$ in IRS is sampled in domain between $\left[ {-\pi/2,\pi/2 } \right]$. Then, the iteration begins and the channel state ${\bf H}_k^0 \in {\Bbb C}^{{M \times 1}}$ is transited into the DNN of policy. Accordingly, the actor part outputs the policy ${\bm \pi}_k^t \left({\bf H}_k^t, {\bm \omega}_k^t\right)$ to realize transmit beamforming. Simultaneously, the critic part outputs Q-value function ${\bf Q}^t\left({\bf H}_k^t, {\bm \upsilon}_k^t\right)$. The choice of ${i_n^t}$ follows the greedy policy. The algorithm selects a random discrete action $i_n$ with the probability $\varepsilon$ and selects ${i_n} = \mathop {\arg \max }\limits_{{{\bf{i}}_n}} {\bf Q}^{t}\left( { {{{\bf H}_k^{t}}},{{\bf{i}}_n}},{{\bm \upsilon}_k^t} \right)$ with probability $1-\varepsilon$. The reward ${\bf r}^t_k = {{{f}}_0^t}\left( {{{\bf{x}}}} \right)$ can be calculated by the obtained channel state ${\bf H}_k^t$ and transmit beamforming ${\bf W}_k^t$. At the same time, the index $i_n^t$ of the IRS derived from the critic part affects the next state ${\bf H}_k^{t+1}$ through new ${\bf V}^{t+1}$.
Therefore, the batch of  $\left\{ {{\bf{H}}_k^t,{{\bm \pi }}_k^t,{\bf{r}}_k^t}, {{\bf H}_k^{t+1}} \right\}$ is obtained. Then batches are collected into the experience memory and parameters of the actor part $\bm \omega$ and the critic part $\bm \upsilon$ can be updated by the experience memory. 
\begin{algorithm}[H]
	\caption{Transmit beamforming and phase-shift design via primal-dual PPO algorithm}
	\begin{algorithmic}
		\STATE  \textbf{Input} Initial transmit beamforming ${\bf W}^0$, policy parameters ${\bm {\omega} ^0}$, Lagrange multipliers ${{\bm \lambda} ^0},{{\bm \upsilon} ^0}$ can be allocated for users. channel states ${\bf H}_0$, ${\bf H}_1$, and ${\bf H}_2$ are fixed in each episode.
		\STATE \textbf{Episode begins}
		\STATE  \textbf{For} $i = 1,2,...,I$ \textbf{do}:
		\STATE \quad \textbf{For} users $ k = 1,2,...,K$ \textbf{do}:
		\STATE  \quad \quad a) Draw samples ${{{\bf \hat x}_1}}$ and ${{{\bf \hat x}_2}}$ from a truncated standard normal distribution;
		\STATE  \quad \quad b) Calculate the channel state ${{\bf H}_k^t} \in {{\Bbb C}^{M \times 1}}$, ${{\bf H}^t = }{{\bf{H}}_0}{\bf{ + }}{{\bf{H}}_2}{{\bf V}^t}{{\bf{H}}_1}$;
		\STATE  \quad \quad c) ${{\bf H}_k^t}$ is fed into the actor part, then the actor part outputs strategy function ${\bf W}_k^t =  {\bm{\pi} _k^t}\left( {{{\bf H}_k^t},{\bm{\omega} _k^t}} \right)$; fed into the critic part, then the critic part outputs Q-value function ${\bf Q}^t\left( { {{\bf H}_k^t},{{\bm \upsilon}_k^t}} \right)$;
		\STATE  \quad \quad d) Calculate the reward ${\bf r}^t_k = {{{f}}_0^t}\left( {{{\bf{x}}}} \right)$ via the formula (23);
		\STATE  \quad \quad e) With the probability $\varepsilon $ select a random discrete action $i_r$ otherwise select ${i_r} = \mathop {\arg \max }\limits_{{{\bf{i}}_r}} {\bf Q}^{t}\left( { {{{\bf H}_k^{t}}},{{\bf{i}}_r}},{{\bm \upsilon}_k^t} \right)$;
		\STATE  \quad \quad f) Calculate the amplitude $\gamma_n$ and phase $\beta_n$ through the formula (3) and (29). Subsequently, the phase-shift design ${\bf V}^{t+1}$ can be obtained;
		\STATE  \quad \quad g) The next state ${{\bf H}_k^{t+1}}$ can be obtained by ${{\bf H}^{t+1} = }{{\bf{H}}_0}{\bf{ + }}{{\bf{H}}_2}{{\bf V}^{t+1}}{{\bf{H}}_1}$;			
		\STATE  \quad \quad h) Collect $\left\{ {{\bf{H}}_k^t,{{\bm \pi }}_k^t,{\bf{r}}_k^t}, {{\bf H}_k^{t+1}} \right\}$ into the experience memory to estimate discounted reward and advantages, then update the actor part and the critic part;
		\STATE  \quad \quad i) Compute the gradient estimate $\widehat \nabla {{{f}}_0}\left( {{{\bf{x}}}} \right)$, $\widehat \nabla {{\bf f}_2}\left( {{{\bf{x}}}} \right)$, ${{{\widehat \nabla }_{\bm{\omega }}}\mathbb{E}\left[ {{\bf{f}_1}\left( {{\bm{\pi }}\left( {{\bf{H}},{\bm{\omega }}} \right),{\bf{h}}} \right)} \right]}$: Execute the formulas (37)--(39);
		\STATE  \quad \quad j) Update the primal-dual variable ${\bm{\omega }}^{t + 1}_k$, ${{\bf{x}}^{t + 1}_k}$, ${{\bm{\lambda}}^{t + 1}_k}$, ${{\bm{\mu}}^{t + 1}_k}$: Execute the formulas (33)--(36);
		\STATE  \quad  Execute step b) to h) until convergence, and an optimized strategy is obtained after the PPO training  ${{\bm \pi}_k ^{\rm{*}}}\left( {{{\bf H}_k},{{\bm \omega}_k}} \right)$;
		\STATE \quad \textbf{End for} 
		\STATE \textbf{End for}
		\STATE  \textbf{Episode ends}
		\STATE  \textbf{Output}  Transmit beamforming ${\bf{W}^{\rm{*}}}$, phase-shift design $\bf{V^{\rm{*}}}$, and system performances ${{f}_0^{\rm{*}}}\left( {{{\bf{x}}}^{\rm{*}}} \right)$.
	\end{algorithmic}
\end{algorithm}
Subsequently, the gradient estimate $\widehat \nabla {{{f}}_0}\left( {{{\bf{x}}}} \right)$, $\widehat \nabla {{\bf f}_2}\left( {{{\bf{x}}}} \right)$, ${{{\widehat \nabla }_{\bm{\omega }}}\mathbb{E}\left[ {{\bf{f}_1}\left( {{\bm{\pi }}\left( {{\bf{H}},{\bm{\omega }}} \right),{\bf{h}}} \right)} \right]}$ can be obtained from the formulas (37)--(39). And primal-dual variables can be obtained through the formulas (33)--(36). After primal-dual PPO training performance converges, the optimized transmit beamforming ${\bf{W}^{\rm{*}}}$, the phase-shift design $\bf{V^{\rm{*}}}$, and the system performances ${{f}_0^{\rm{*}}}\left( {{{\bf{x}}}^{\rm{*}}} \right)$ can be achieved. For the sake of clarity, a full notation list is included in TABLE I.
\begin{table}
	\caption{List of Notations.}
	\begin{center}
		\setlength{\tabcolsep}{1mm}{
			\begin{tabular}{|c|c|}
				\hline
				{\bf Notations} & { \bf Explanations}\\
				\hline
				$\cal M$ & {The set of ISAC BS's antennas} \\
				\hline
				$\cal K$ & {The set of served users}\\
				\hline
				$\cal N$ & {The set of IRSs' elements}\\
				\hline
				$\cal R$ & {The set of user's antennas}\\
				\hline
				$\bf A$ & {The target response matrix of the ISAC BS}\\
				\hline
				$\bf B$ & {The target response matrix of the IRS's elements}\\
				\hline
				${\bf V}$ & {The effective diagonal phase matrix}\\
				\hline
				$i_n$& {The discrete action of the IRS's element index}\\
				\hline
				$\gamma$ & {The amplitude reflection coefficient of the IRS}\\
				\hline
				$\beta$ & {The phase-shift coefficient of the IRS} \\
				\hline
				$\theta$ & {The azimuth angle}\\
				\hline
				$\theta_{AoD}$ & {The angle of departure}\\
				\hline
				$\theta_{AoA}$ & {The angle of arrival}\\
				\hline
				$\psi_l$ & {The sampled angle}\\	
				\hline
				$d\left( {{\psi}} \right)$ & {The desired beam pattern in direction $\psi$}\\
				\hline
				$d\left( {{\psi_p}} \right)$ & {The ideal beam pattern in direction $\psi_p$}\\
				\hline
				$\phi$ & {The elevation angle}\\				
				\hline
								${\bf{a}}\left( {{\theta}} \right)$ & {Steering vector at the ISAC BS}\\
				\hline
				${{\mathbf{v}}}\left( {\phi ,{\mathbf{\theta }}} \right)$&{Steering vector at the IRSs}\\
				\hline
				${\bf{b}}\left( {{\theta}} \right)$ & {Steering vector at receive users}\\
				\hline
				$P\left( \psi  \right)$ & {The power consumption in direction $\psi$}\\
				\hline
				$P_{max}$ & {The total transmit power}\\
				\hline
				$\bf R$&{The covariance matrix}\\
				\hline
				${\bf H}_0$ & {The channel gain from the ISAC BS to user $k$}\\
				\hline
				${\bf H}_1$ & {The channel gain from the ISAC BS to the IRS}\\
				\hline
				${\bf H}_2$ & {The channel gain from the IRS's element $n$ to user $k$}\\
				\hline
				${\bf H}_{LOS}$ & {The LoS component}\\
				\hline
				${\bf H}_{NLOS}$ & {The  NLoS component}\\
				\hline
				$\bf x$&The ergodic average value\\
				\hline
				${{f}_0}\left(  \cdot  \right)$ & {The sum function}\\
				\hline
				${{\bf f}_1}\left(  \cdot  \right)$& {The instantaneous performance function}\\
				\hline
				${{\bf f}_2}\left(  \cdot  \right)$& {The constraint function}\\
				\hline
				$\bm \omega$& {The parameters of the actor neural network}\\
				\hline
				$\bm \upsilon $& {The parameters of the critic neural network}\\
				\hline
				$\ell $&The threshold at the target\\	
				\hline
				${\bf{A}}\left( {{\bf{H},\bm{\omega }}} \right)$& The advantage function\\
				\hline
				$ {v}\left( {{{\bf{H}},{\bm \upsilon}}} \right)$&{The value function of critic part}\\
				\hline
				${\bf Q}\left( { {{{\bf H}}},{{\bf{i}}}},{{\bm \upsilon}} \right)$&The Q-value function\\
				\hline
				${\bm \pi} \left({\bf H}, {\bm \omega}\right)$& The policy function in the current episode\\
				\hline
				${\bm \pi}_{old} \left({\bf H}, {\bm \omega}\right)$& The policy function in the former episode\\
				\hline
				${L^{PG}}\left( {\bm{\omega }} \right)$&{The loss function of policy gradient method}\\
				\hline
				${L^{PPO}}\left( {\bm{\omega }} \right)$&{The loss function of PPO method} \\
				\hline
				${L^{CLIP}}\left( {\bm{\omega }} \right)$&{The loss function of simplified PPO method}\\
				\hline
				$\bm \lambda$, $\bm{\mu }$& {The Lagrange multipliers}\\

				\hline
				$\tau$& The size of updating step length\\
				\hline
				$\alpha_1$, $\alpha_2$, $\alpha_3$&{The gradient estumated upgrading step factors}\\
				\hline
				$\epsilon$ & {The hyperparameter of PPO}\\											
				\hline
				$\varepsilon$ & {The exploring rate of greedy policy of DQN }\\
				\hline
		\end{tabular}}
	\end{center}
	\label{table:1}
\end{table}
\subsection{Primal-dual DPPO Algorithm for Multi-user MIMO Scenario}
\begin{algorithm}
	\caption{Transmit beamforming and phase-shift design via primal-dual DPPO algorithm}
	\begin{algorithmic}
		\STATE \textbf{Input} Initial beamforming $\bf {W}^0$,  Lagrange multipliers ${\bm \lambda}^0,{\bm \mu}^0$, policy parameters ${{\bm \omega}^0}$ and DQN's parameters ${{\bm \upsilon} ^0}$. Channel states ${\bf H}_{0,k}$, ${\bf H}_{1,k}$, and ${\bf H}_{2,k}$ are fixed in each episode.
		\STATE  \textbf{Episode begins}
		\STATE  \textbf{For} $i = 1,2,...,I$ \textbf{do}:
		\STATE \quad Wait until a certain amount of gradients $\bm \omega$ are available average gradients and update global $\bm \omega$. Wait until a certain amount of $\bm \upsilon$ are available average gradients and update global $\bm \upsilon$ (workers stop collecting data when the chief is updating).		
	    \STATE  \quad \textbf{For} the worker (user) $k = 1,2,...,K$ \textbf{do}:
		\STATE  \quad \quad \textbf{For} received antennas $ n = 1,2,...,N$ \textbf{do}:
		\STATE  \quad \quad \quad a) Draw samples ${{{\bf \hat x}_1}}$ and ${{{\bf \hat x}_2}}$ from a truncated standard normal distribution;
		\STATE  \quad \quad \quad b) Channel state ${{\bf H}_n^t} \in {{\Bbb C}^{M \times 1}}$, obtained from the matrix of ${{\bf H}^t = }{{\bf{H}}_0}{\bf{ + }}{{\bf{H}}_2}{{\bf V}^t}{{\bf{H}}_1}$;
		\STATE  \quad \quad \quad c) Calculate the reward ${\bf r}^t_k = {{{f}}_0^t}\left( {{{\bf{x}}}} \right)$ via the formula (23);
		\STATE  \quad \quad \quad d) Each worker $k$ repeats step d) to step j) in Algorithm 1;
		\STATE  \quad \quad \textbf{End for}
		\STATE \quad  \textbf{Worker (user) ends}
		\STATE \textbf{End for}
		\STATE  \textbf{Episode ends}
		\STATE  \textbf{Output} Transmit beamforming ${\bf{W}^{\rm{*}}}$, phase-shift design $\bf{V^{\rm{*}}}$, and the system performances ${{f}_0^{\rm{*}}}\left( {{{\bf{x}}}^{\rm{*}}} \right)$.
	\end{algorithmic}
\end{algorithm}
For multi-user MIMO scenario, the pseudo code of the proposed primal-dual DPPO is concluded in Algorithm 2. As discussed in Section IV. B, each worker takes charge of each user. They transmit respective gradients of DNNs to the global PPO. The global PPO updates policy parameters ${{\bm \omega}^0}$ and DQN's parameters ${{\bm \upsilon} ^0}$ based on the estimated advantage functions ${\hat {\bf A}}^t\left({\bf H},{\bm \omega}\right)$. Each worker (i.e. user) utilizes the original primal-dual PPO algorithm in Algorithm 1. 
\section{Numerical Simulation and Analysis}
The simulation results in this section. We simulated in an x64 workstation with Intel i7 CPU and the Microsoft Windows 10 Operation System. The version of Python and TensorFlow is v3.7.9 and v2.1.0, respectively. According to the work in \cite{Reference36}, the carrier frequency is set as 0.55 THz and the factor of molecular absorption loss ${k_{abs}}$ is $6.7141 \times {10^{ - 4}}$. 
The paper simulated IRS-aided ISAC system where several users are randomly distributed in need. The targets' azimuth angles $\theta_k$ are assumed in ${\text{-40}}^\circ$, ${\text{0}}^\circ$, and ${\text{40}}^\circ$. The $\Delta$ in the formula (17) is set as ${\text{5}}^\circ$ and the ideal beam pattern $\theta_p$ consists of three directions, including $\theta_1 = -40^\circ$, $\theta_2 = 0^\circ$, and $\theta_3 = 40^\circ$ \cite{Reference37}.

The PPO's structure is divided into two DNNs, including the actor part and the critic part. The primal-dual PPO optimization is performed with Adam optimizer for these two DNNs parameters update.
The layer of actor part is a single dense layer with 30 units and its activation function is ReLU. The policy ${\bm \pi}_{\omega}$ is derived from the truncated Gaussian distribution. In the truncated Gaussian distribution, the mean is the output of dense layer with tanh activation function and the standard deviation is the output of dense layer with softplus activation function. The action of policy ${\bf W}_k \in {{\Bbb C}^{M \times 1}} $ is fixed on the domain $\left[0, P_{max}\right]$. The maximum time step in one episode is set as 50.

There exists a single dense layer with 20 units in the critic part. And the activation function of the dense layer is ReLU. Besides output of value function, the critic part also outputs the Q-value function. It has the function of adjusting the IRS'elements and it can generate the indexes of phases shift. The action choice of Q-value function follows $\varepsilon\text{-greedy}$ strategy. The parameter of $\varepsilon\text{-greedy}$ strategy is set as $\varepsilon=0.95$. Thus, the critic part outputs two values, including  the baseline function \cite{Reference35} and Q-value function. For updating the primal and dual variables, the batch size of replay experience is set to 32. And the primal dual upgrade steps are set as ${\gamma _{\rm{1}}}{\rm{ = }}{\gamma _{\rm{2}}}{\rm{ = }}{\gamma _{\rm{3}}}{\rm{ = }}{\gamma _{\rm{4}}}{\rm{ = 0}}{\rm{.001}}$, and the gradient upgrades are set as ${\alpha _{\rm{1}}}{\rm{ = }}{\alpha _{\rm{2}}}{\rm{ = }}{\alpha _{\rm{3}}}{\rm{ = }}{\rm{  0}}{\rm{.001}}$.

\subsection{Multi-user MISO Scenario}
In the simulation of multi-user MISO scenario, we deploy a ISAC BS equipped with 5 antennas and 4 users with single antenna in IRS-aided ISAC system. In this case, the proposed primal-dual PPO algorithm obtains each user's channel state ${\bf H}_k \in {{\Bbb C}^{M \times 1}}$ in turn within each episode.

\begin{figure}
	\centering
	\includegraphics[width=85mm]{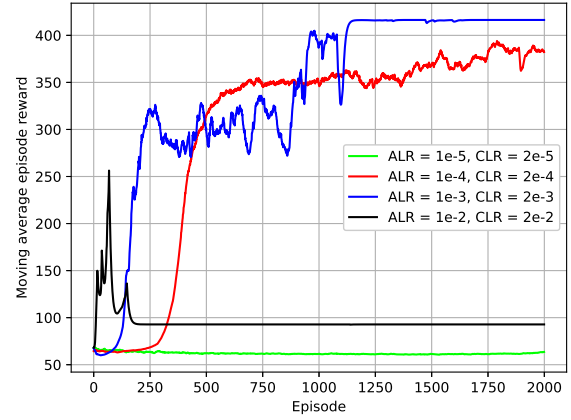}
	\caption{The reward of primal-dual PPO algorithm with different learning rates.}
	\label{fig3}
\end{figure}

In Fig. 3, the reward of the proposed primal-dual PPO is simulated in different learning rates. The appropriate learning rate can avoid unnecessary training and achieve quicker optimization. As shown in Fig. 3, when the learning rate of actor equals to ${\text{1}} \times {\text{1}}{{\text{0}}^{{\text{-3}}}}$ and the learning rate of critic equals to ${\text{2}} \times {\text{1}}{{\text{0}}^{{\text{-3}}}}$, the moving averaged episode reward can convergence after 1200 episodes. However, when the order of magnitude is $-\text{2}$ in learning rate (the learning rate of the actor part is ${\text{1}} \times {\text{1}}{{\text{0}}^{{\text{-2}}}}$ and the learning rate of the critic part is ${\text{2}} \times {\text{1}}{{\text{0}}^{{\text{-2}}}}$), it remains in around 90. Similarly, $-\text{5}$ order of magnitude (the learning rate of the actor part is ${\text{1}} \times {\text{1}}{{\text{0}}^{{\text{-5}}}}$ and the learning rate of the critic part is ${\text{2}} \times {\text{1}}{{\text{0}}^{{\text{-5}}}}$) also keeps in around 70 due to too small learning rate. Although $-\text{4}$ order of magnitude (the learning rate of the actor part is \ ${\text{1}} \times {\text{1}}{{\text{0}}^{{\text{-4}}}}$ and the learning rate of the critic part is ${\text{2}} \times {\text{1}}{{\text{0}}^{{\text{-4}}}}$) has tendency to converge, it still requires more unnecessary episodes to converge. The step size of learning rate is so big that the agent does not learn the suitable action. Using so many learning parameters, it can be seen that the learning rate of actor equals to ${\text{1}} \times {\text{1}}{{\text{0}}^{{\text{-3}}}}$ and the learning rate of critic equals to ${\text{2}} \times {\text{1}}{{\text{0}}^{{\text{-3}}}}$, the moving averaged episode reward can convergence after 1200 episodes. Too high or too small learning rate is not suitable in the designed model.
\begin{figure}
	\centering
	\includegraphics[width=85mm]{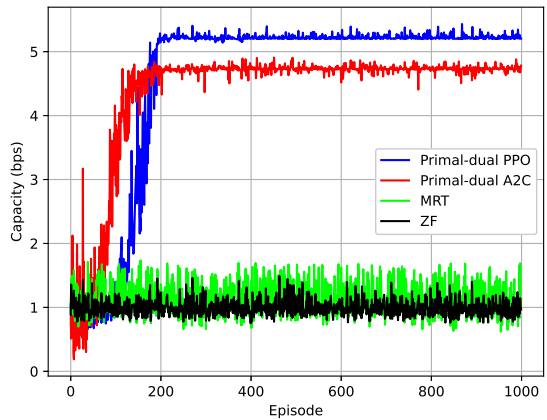}
	\caption{Comparison of performance using primal-dual PPO algorithm and other algorithms.}
	\label{fig4}		
\end{figure}
Fig. 4 shows the capacity of IRS-aided ISAC system based on different algorithms. The zero force (ZF) transmit beamforming and maximum ratio transmission (MRT) are introduced to compare with the proposed primal dual PPO algorithm \cite{Reference38}. Initially, the performance function $f_0\left({\bf x}\right)$ (i.e. total capacity) with the primal-dual PPO algorithm is lower than the previous two methods slightly. Nevertheless, the obtained system capacity converges after 200 episodes. Due to the shifting environment in each episode, it can be seen that ZF and MRT can only obtain around 1 bps capacity in the same training conditions. These traditional optimization occupy too high computational cost so that is not suitable the practical IRS-aided ISAC system. Referring to the primal-dual PPO, a primal-dual advantage actor-critic (A2C) algorithm is designed. Although primal-dual A2C can obtain the similar convergence, there still exists around 0.5 bps gap between two different primal-dual learning algorithms. This is because PPO's important sampling ensures new policy can not deviate too far. The new improved policy can be modified by the old policy constantly. Compared with the primal-dual A2C algorithm, the proposed primal-dual PPO method is more efficient and robust.

\begin{figure}
	\centering
	\includegraphics[width=85mm]{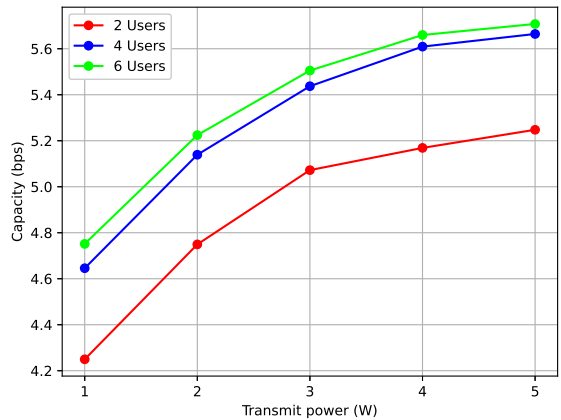}
	\caption{Comparison of capacity versus transmit power under different users.}
	\label{fig5}
\end{figure}
Fig. 5 studies the capacity comparison versus different transmit powers for different users. It is clear that as the ISAC BS transmit power increases, the performance function increases as well. It reflects that more transmit power contributes to realizing higher capacity of IRS-aided ISAC system. Meanwhile, the gap between different users does not change significantly. The simulation result reveals the more receive users obtain higher capacity on the condition that equal transmit power. When transmit power is big enough, the increasing speed of capacity tends to slow. The reason is that the logarithmic sum relationship between ${f_0}\left( {{{\mathbf{x}}}} \right)$ and policy ${\bm{\pi }}\left( {{\bf{H}},{\bm{\omega }}} \right)$.

\begin{figure}
	\centering
	\includegraphics[width=85mm]{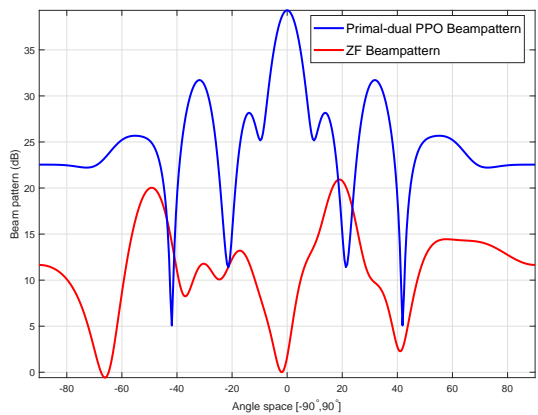}
	\caption{Comparison of beam patterns using primal-dual PPO algorithm and ZF algorithm.}
	\label{fig6}
	
\end{figure}
Fig. 6 shows the beam pattern is optimized by the primal-dual PPO algorithm and the ZF algorithm. 
The transmit beam pattern is calculated by the covariance of transmit waveform $\bf R$.
It shows beam pattern against different angle spaces, where the angle ranges from ${- \rm{90}}^\circ$ to ${\rm{90}}^\circ$. It can be seen that the performance of detecting targets is focused on the desired angles, including ${-\rm{40}}^\circ$, ${\rm{0}}^\circ$, and ${\rm{40}}^\circ$. Compared to the ZF beam pattern, the beam pattern optimized by the primal-dual PPO algorithm has a better detecting performance in the proposed multi-user MISO scenario.	
\begin{figure}
	\centering
	\includegraphics[width=85mm]{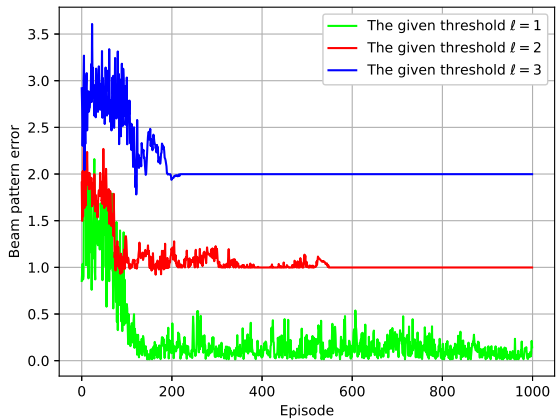}
	\caption{Beam pattern error versus different thresholds $\varepsilon $.}
	\label{fig7}
\end{figure}

Fig. 7 shows the beam pattern error $L_r\left({\bf R}\right)$ under different thresholds $\ell$. The different thresholds are compared among 1, 2, and 3. It can be seen that the lower threshold $\ell$ can obtain the lower beam pattern error $L_r\left({\bf R}\right)$. When $L_r\left({\bf R}\right) \le 1$, the beam pattern error $L_r\left({\bf R}\right)$ fluctuates among 0.2. However, the lower threshold $\ell$, the learning time requested is longer. This is because the higher sensing accuracy need longer time to learning.

\begin{figure}
	\centering
	\includegraphics[width=85mm]{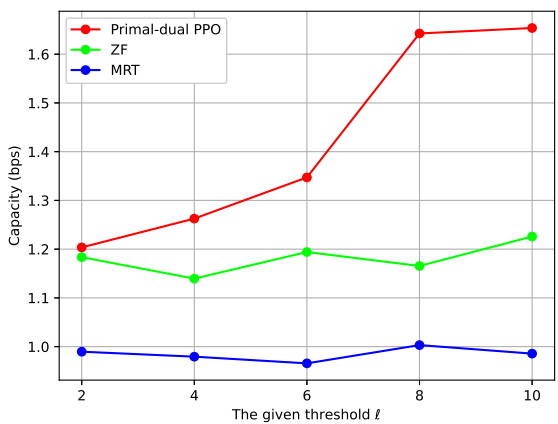}
	\caption{Radar-communication trade-off performance.}
	\label{fig1}
\end{figure}

Fig. 8 indicates the radar-communication trade-off performance in the proposed ISAC system. The different thresholds are increasing form 2 to 10. It is obvious that the larger given threshold, the obtained system capacity will be larger in the proposed algorithm. This is because the tolerance of radar performance, MSE, relieve the pressure from sensing function. The higher communication performance will be obtained when $ \ell$ equals 10.

\begin{figure}[H]
	\centering
	\includegraphics[width=85mm]{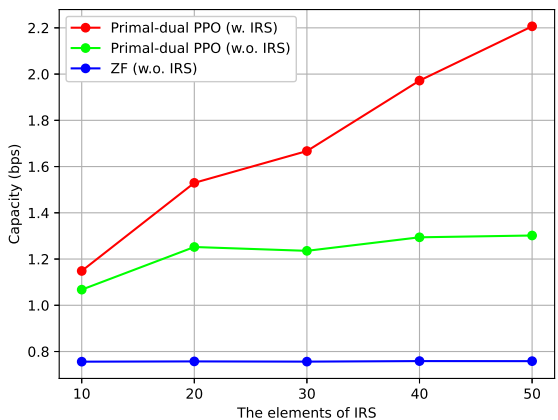}
	\caption{System capacity versus the number of reflective elements.}
	\label{fig7}
\end{figure}

Fig. 9 reveals the system capacity versus the number of reflective elements. We compared with two different methods, including with IRS relaying and without IRS relaying. It can be seen that the system capacity increases with the increasing of the elements of IRS, under the condition of IRS relaying. Obviously, when the IRS is not aided in the ISAC system, the capacity does not change. It suggests the function of IRS is significant in the field of enhancing capacity.

\subsection{Multi-user MIMO Scenario}
After discussing the simulation results in multi-user MISO scenario in IRS-aided ISAC system. The figures about multi-user MIMO scenario are shown in this subsection. The number of the transmit antennas of the ISAC BS is set as 5 and the received antennas of each user is set as 4. The number of DPPO's workers equals to the number of severed users. The learning rate of the actor part and the critic part is $1 \times 10^{-3}$ and $2 \times 10^{-3}$, respectively. The primal dual upgrade steps $\gamma_1$, $\gamma_2$, $\gamma_3$, $\gamma_4$ and gradient upgrades $\alpha_1$, $\alpha_2$, $\alpha_3$ are the same as the case of multi-user MISO scenario.

Fig. 10 shows that the moving averaged episode reward with the growing of episode. To verify the effectiveness of the proposed primal-dual DPPO algorithm, A3C algorithm and weighted minimum mean squared error (WMMSE) algorithm \cite{Reference29} are introduced in this simulation. More detailed discussion about these two algorithms are introduced in the Appendix B. As shown in Fig. 10, the A3C algorithm also has a tendency to increase, but the speed of learning is too slow under the same learning rate. This is the reason that the learning rate is not suitable in A3C method, resulting in the learning effect is slow. Meanwhile, we also introduce the WMMSE algorithm as an unparameterized method and calculate its reward in each episode. Obviously, the proposed primal-dual DPPO converges after 800 episodes, and achieves a higher reward than two other algorithms'.
\begin{figure}[H]
	\centering
	\includegraphics[width=85mm]{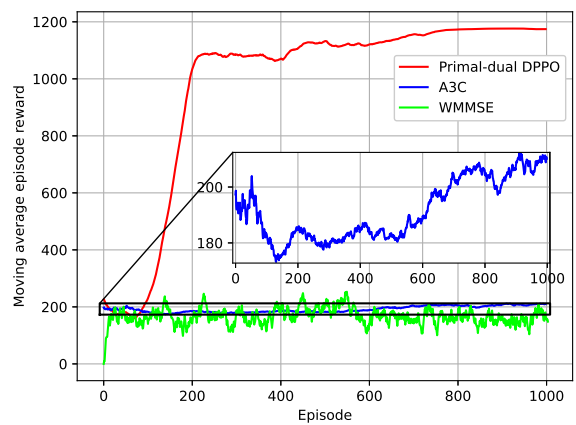}
	\caption{Convergence of moving averaged episode reward using different algorithms.}
	\label{fig8}
\end{figure}

Fig. 11 compares the capacity of different distributed DRL algorithms. Obviously, the proposed primal-dual DPPO algorithm can converge after 90 episodes and it shows more stable performance. This is because the global PPO collects local data rather than policy gradients from different workers. Compared with the parallel optimization in the A3C algorithm, the primal-dual DPPO algorithm can save computation time and realize quicker optimization.
\begin{figure}[H]
	\centering
	\includegraphics[width=85mm]{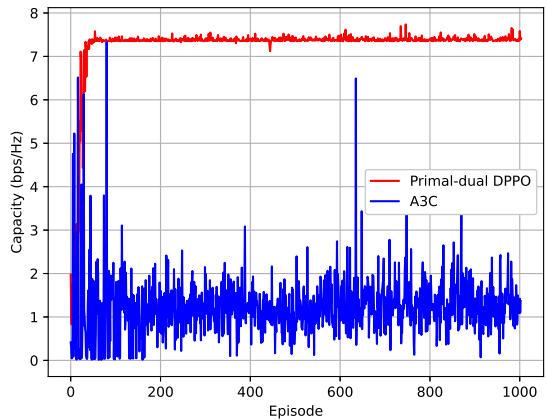}
	\caption{Comparison of performance using primal-dual DPPO algorithm and A3C algorithm.}
	\label{fig9}
\end{figure}
Fig. 12 illustrates the beam patterns of 4 users in the multi-user MIMO scenario. The number of transmit antennas is set as 5. The targets are also focused on the desired angles, including ${-\rm{40}}^\circ$, ${\rm{0}}^\circ$, and ${\rm{40}}^\circ$. It can be seen that each user realizes the desired performance of detecting targets via the proposed primal-dual DPPO algorithm. The amplitudes of 4 users may be different. The reason is that the different users may face with different channel states observations. Different workers take charges of different users, it implies that the utilization rate of process is fully scheduled.

\begin{figure}[H]
	\centering
	\includegraphics[width=85mm]{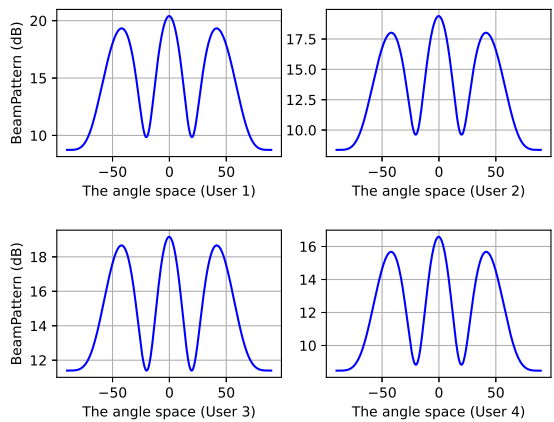}
	\caption{Beam patterns in different users via primal-dual DPPO algorithm.}
	\label{fig10}
\end{figure}
\section{Conclusion}
In this paper, we have investigated capacity maximization in the IRS-aided ISAC system and considered the joint transmit beamforming and passive phase optimization. The optimization problem was transformed into an ergodic form to capture the long-term ISAC system performance. Considering the power budget and target response, a beamforming optimization scheme combining PPO with primal-dual was adopted in the multi-user MISO-based scenario. Furthermore, we proposed to utilize the concept of multi-threading derived from the DPPO, to realize the joint optimization of transmit beamforming and phase-shift design in the multi-user MIMO scenario. Simulations results have verified the effectiveness of the primal-dual PPO algorithm and the primal-dual DPPO algorithm. With the aid of the actor-critic structure in DRL, the joint optimization transmit beamforming and phase-shift design achieved higher capacity.  
\appendices
\section{Proof of the Trust Region Policy Optimization}
The goal of intensive learning is to obtain the best strategy that maximizes the desired reward function.
\begin{equation}
	\eta \left( {{{\bm{\pi }}^{t+1}}} \right) =  {\mathbb{E}_{{{\mathbf{H}}_0},{p_0},...,{{\bm{\pi }}^t}}}\left[ {\sum\limits_{t = 0}^\infty  {{\gamma ^t}{{\bf A}^t}\left( {{{\mathbf{H}}^t},{\bm{\omega }}} \right)} } \right]
\end{equation}

The discounted visitation frequencies ${\rho _{\bm \pi} }\left( {\mathbf{H}} \right) = \sum\limits_{t = 0}^\infty  {{\gamma ^t}P\left( {{{\mathbf{H}}^t} = {\mathbf{H}}} \right)} $ are introduced to the formula (45), it can be transformed into 
\begin{equation}
	\eta \left( {{{\bm{\pi }}^{t+1}}} \right)  = \sum\limits_{\mathbf{H}} {{\rho _{{{\bm{\pi }}^{t + 1}}}}\left( {\mathbf{H}} \right)} \sum\limits_p {{{\bm{\pi }}^{t + 1}}\left( {p\left| {\mathbf{H}} \right.} \right)} {{\bf A}^t}\left( {{{\mathbf{H}}^t},{\bm{\omega }}} \right)
\end{equation}
However, the discounted visitation frequencies ${\rho _{{\bm \pi}}}\left( {\mathbf{H}} \right)$ of new policy ${\bm \pi}^{t+1}$ in the formula (46) has to be calculated, which requires too high computation cost. The surrogate function ${L_{\bm{\pi }}}\left( {{{\bm{\pi }}^{t + 1}}} \right)$ is introduced to approximate the $\eta \left( {{{\bm{\pi }}^{t + 1}}} \right)$:
\begin{equation}
	{L_{\bm{\pi }}}\left( {{{\bm{\pi }}^{t + 1}}} \right) = \sum\limits_{\mathbf{H}} {{\rho _{\bm{\pi }}}\left( {\mathbf{H}} \right)} \sum\limits_p {{{\bm{\pi }}^{t + 1}}\left( {p\left| {\mathbf{H}} \right.} \right)} {{\bf A}^t}\left( {{{\mathbf{H}}^t},{\bm{\omega }}} \right)
\end{equation}
The difference between ${L_{\bm{\pi }}}\left( {{{\bm{\pi }}^{t + 1}}} \right)$ and $\eta \left( {{{\bm{\pi }}^{t + 1}}} \right)$ is different ${\rho _{\bm{\pi }}}\left( {\mathbf{H}} \right)$. The discounted visitation frequencies ${\rho _{\bm \pi}}\left( {\mathbf{H}} \right)$ of the surrogate function ${L_{\bm{\pi }}}\left( {{{\bm{\pi }}^{t + 1}}} \right)$ can be obtained by the old policy ${\bm \pi}^{t}$. When ${\bm \pi}^{t}$ and ${\bm \pi}^{t+1}$ satisfy the constraints, they can be equal.

Herein, the total variation (TV) divergence $D_{TV}$ is utilized to deign the constraints. For two discrete probabilities $p$ and $q$, the TV divergence ${D_{TV}}\left( {p\left\| q \right.} \right) = \frac{1}{2}\sum\limits_i {\left| {{p_i} - {q_i}} \right|} $. It is assumed that $D_{TV}^{\max }\left( {{{\mathbf{\pi }}^t},{{\mathbf{\pi }}^{t + 1}}} \right) = \mathop {\max }\limits_{\mathbf{H}} {D_{TV}}\left( {{{{\bm \pi }}^t}\left( {\mathbf{H}} \right)\left\| {{{{\bm \pi }}^{t + 1}}\left( {\mathbf{H}} \right)} \right.} \right)$, the difference of the surrogate function ${L_{\bm{\pi }}}\left( {{{\bm{\pi }}^{t + 1}}} \right)$ and the cumulative reward $\eta \left( {\bm{\pi }} \right)$ is given by
\begin{equation}
	\eta \left( {{\bm \pi }} \right) \geqslant {L_\pi }\left( {{\bm \pi }} \right) - \frac{{4\varepsilon \gamma }}{{{{\left( {1 - \gamma } \right)}^2}}}D_{TV}^{\max }\left( {{{{\bm \pi }}^t},{{{\bm \pi }}^{t + 1}}} \right)
\end{equation}

The relationship between TV divergence and KL divergence satisfy ${D_{TV}}{\left( {p\left\| q \right.} \right)^2} \leqslant {D_{KL}}\left( {p\left\| q \right.} \right)$, the constraint (48) can be reformulated by the DL divergence.
\begin{equation}
	\eta \left( {{{{\bm \pi }}^{t + 1}}} \right) \geqslant {L_\pi }\left( {{{{\bm \pi }}^{t + 1}}} \right) - \frac{{4\varepsilon \gamma }}{{{{\left( {1 - \gamma } \right)}^2}}}D_{KL}^{\max }\left( {{{{\bm \pi }}^t},{{{\bm \pi }}^{t + 1}}} \right)
\end{equation}
With the constraint of the KL divergence, the lower bound of the cumulative reward $\eta \left( {\bm{\pi }} \right)$ can be obtained. Additionally, the true objective $\eta\left({\bm \pi}\right)$ is non-decreasing. The following constraint also can guarantee the optimization of the cumulative reward $\eta \left( {\bm{\pi }} \right)$. 

\begin{equation}
	\mathop {\max }\limits_\theta  {L_{{{\mathbf{\pi }}_{{\theta _0}}}}}\left( {{{\mathbf{\pi }}_\theta }} \right) - \frac{{4\varepsilon \gamma }}{{{{\left( {1 - \gamma } \right)}^2}}}D_{KL}^{\max }\left( {{{\mathbf{\pi }}_\theta },{{\mathbf{\pi }}_{{\theta _0}}}} \right)
\end{equation}
In practical, the penalty coefficient $\frac{{4\varepsilon \gamma }}{{{{\left( {1 - \gamma } \right)}^2}}}$ will cause the step length of policy iteration small. One way to take larger steps is to use a constraint based on the KL divergence between the new policy and the old policy, i.e., a trust region constraint.
\begin{equation}
	\begin{array}{*{20}{l}}
		{\mathop {\max }\limits_{\bm{\omega }} }&{{\mathbb{E}_{{\rho _\omega }\left( \tau  \right)}}\left[ {\sum\nolimits_t {{\gamma ^{t - 1}}\frac{{{\bm{\pi }}\left( {{\mathbf{H}},{\bm{\omega }}} \right)}}{{{{\bm{\pi }}_{old}}\left( {{\mathbf{H}},{\bm{\omega }}} \right)}}{{\mathbf{A}}_{old}}\left( {{\mathbf{H}},{\bm{\omega }}} \right)} } \right]}, \\ 
		{s.t.}&{{D_{KL}}\left[ {\frac{{{\bm{\pi }}\left( {{\mathbf{H}},{\bm{\omega }}} \right)}}{{{{\bm{\pi }}_{old}}\left( {{\mathbf{H}},{\bm{\omega }}} \right)}}} \right] \leqslant \delta } .
	\end{array}
\end{equation}

\section{Introduction of A3C algorithm and WMMSE algorithm}
\subsection{A3C Algorithm}
The A3C algorithm is one of the methods for asynchronous DRL in a multi-threading mode. 
The A3C algorithm essentially puts actor-critic algorithm into multiple threads for synchronous training.
On the one hand, it is based on the policy-based method, so it can deal with continuous state and action spaces. On the other hand, A3C algorithm employs asynchronous method to relieve data correlation, in which data is not generated at the same time. Compared with the DQN algorithm, the A3C algorithm does not need to use the experience pool to store historical samples. This asynchronous method can save storage space and accelerate the sampling speed of data. All the actor learners update the policy ${\bm{\pi }}\left( {{\mathbf{H}},{\bm{\omega }}} \right)$
and the advantage function ${\mathbf{A}}\left( {{\mathbf{H}},{\bm{\omega }}} \right)$ according to gradient loss. Based on the advantage function, the gradient loss function is given by
\begin{equation}
	{L^{A3C}}\left( {\bm {\omega }} \right){\text{ = }}\mathbb{E}\left[ {{\nabla _{\bm{\omega }}}\log {\bm{\pi }}\left( {{\mathbf{H}},{\bm{\omega }}} \right){\mathbf{A}}\left( {{\mathbf{H}},{\bm{\omega }}} \right)} \right]
\end{equation}
The update formulas for $\bm \omega$ and $\bm \upsilon$ are
\begin{equation}
	\begin{array}{*{20}{l}}
		{{{\bm{\omega }}^{t + 1}} = {{\bm{\omega }}^t} + {\nabla _{\bm{\omega }}}\log {\bm{\pi }}\left( {{\mathbf{H}},{\bm{\omega }}} \right)\left( {r - v\left( {{\mathbf{H}},{{\bm \upsilon} ^t}} \right)} \right)} \\ 
		{{{\bm \upsilon} ^{t + 1}} = {{\bm \upsilon} ^t} + \partial {{\left( {r - v\left( {{\mathbf{H}},{{\bm \upsilon} ^t}} \right)} \right)} \mathord{\left/
					{\vphantom {{\left( {r - v\left( {{\mathbf{H}},{{\bm \upsilon} ^t}} \right)} \right)} {\partial {\bm \upsilon} }}} \right.
					\kern-\nulldelimiterspace} {\partial {\bm \upsilon} }}} 
	\end{array}
\end{equation}

\subsection{WMMSE Algorithm}
This algorithm transforms the weighted sum-rate maximization problem into a higher dimensional space, using the well-known minimum mean squared error. The sum-rate maximization is equivalent to the following weighted sum-MSE minimization,
\begin{equation}
	\begin{array}{*{20}{l}}
		{\mathop {\min }\limits_{\varsigma ,\vartheta , W } }&{\sum\limits_{k = 1}^K {\left( {{\varsigma _k}{e_k} - \log \left( {{\varsigma _k}} \right)} \right)} } \\ 
		{s.t.}&{0 \leqslant {W_k} \leqslant \sqrt {{P_{\max }}} ,k = 1,2,...,K.} 
	\end{array}
\end{equation}
where $e_k$ is the meansquare estimation error and $\vartheta_k$ equals the ${SINR}_k$.
\begin{equation}
	{e_k} = {\left( {{\vartheta _k}{{\mathbf{H}}_k}{{\mathbf{W}}_k} - 1} \right)^2} + \sum\limits_{l \ne k} {{{\left( {{\vartheta _k}{{\mathbf{H}}_l}{{\mathbf{W}}_l}} \right)}^2} + \sigma _c^2\vartheta _k^2} 
\end{equation}
The $\varsigma_k = e_k^{-1}$. Here by equivalent we meant that all stationary solutions of the sum-MSE minimization is identical to maximizing sum rate. 

%



%

\begin{IEEEbiography}[{\includegraphics[width=1in,height=1.25in,clip,keepaspectratio]{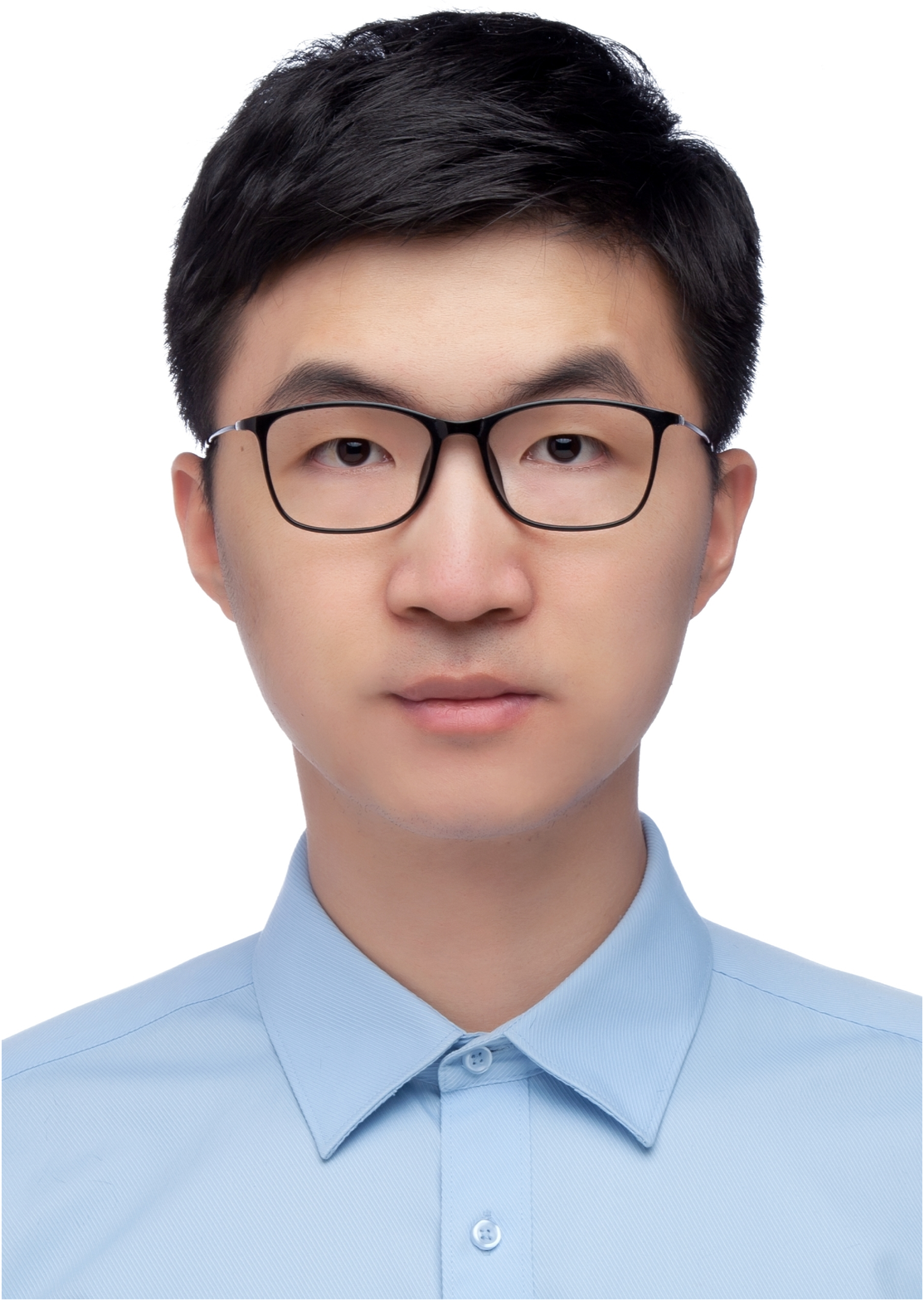}}]{Xiangnan Liu} received the B.S. degree from the School of Computer and Communication Engineering, University of Science and Technology of Beijing, Beijing, China, in 2019. He is currently pursuing his Ph.D. degree at University of Science and Technology Beijing, China. His research interests include access control, beamforming, and resource allocation in 6G wireless communication.	
\end{IEEEbiography}
\begin{IEEEbiography}[{\includegraphics[width=1in,height=1.25in,clip,keepaspectratio]{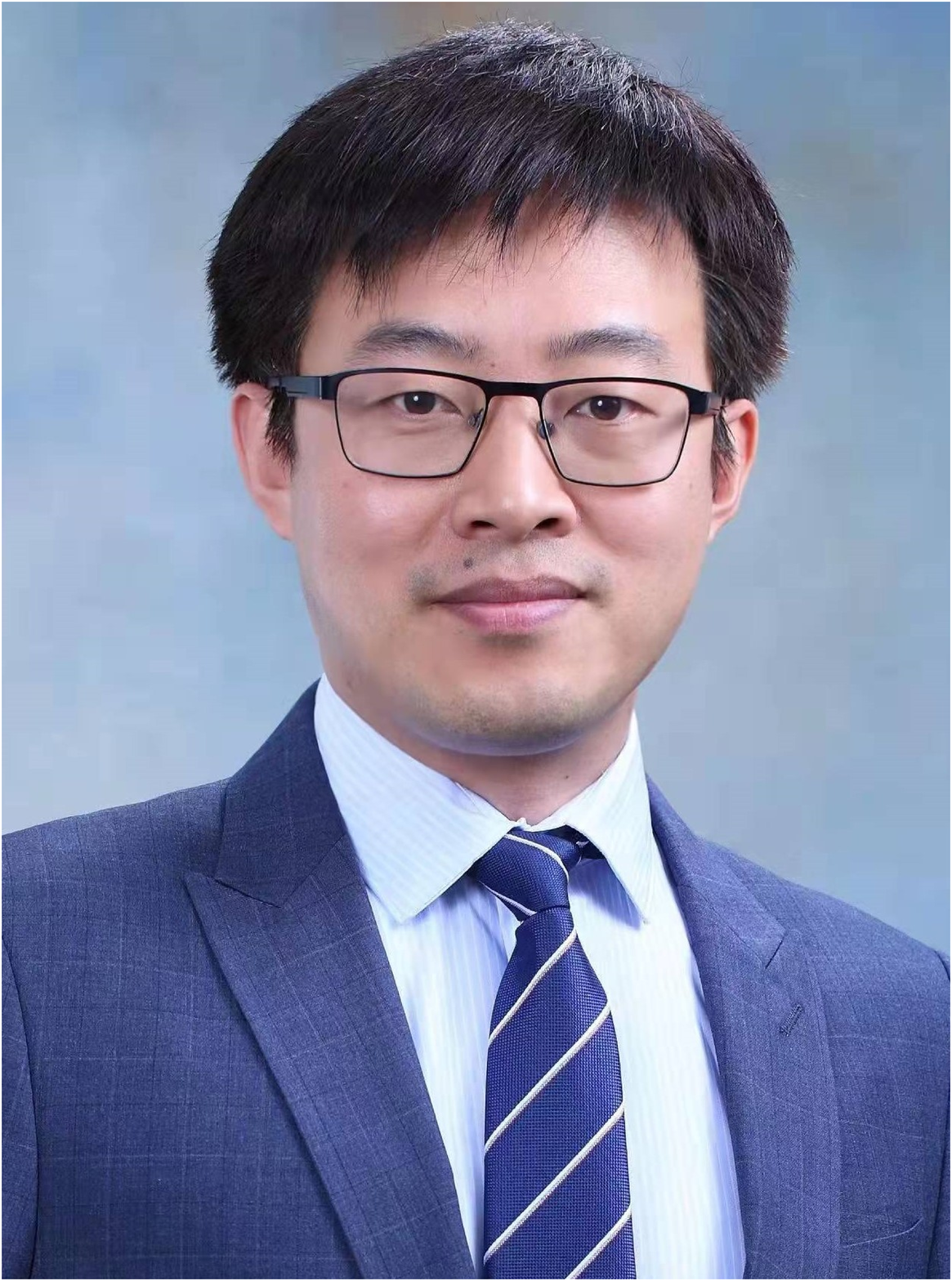}}]{Haijun Zhang}
	(M’13, SM’17) is currently a Full Professor and Associate Dean at University of Science and Technology Beijing, China. He was a Postdoctoral Research Fellow in Department of Electrical and Computer Engineering, the University of British Columbia (UBC), Canada. He serves/served as Track Co-Chair of WCNC 2020, Symposium Chair of Globecom'19, TPC Co-Chair of INFOCOM 2018 Workshop on Integrating Edge Computing, Caching, and Offloading in Next Generation Networks, and General Co-Chair of GameNets'16. He serves as an Editor of IEEE Transactions on Communications, IEEE Transactions on Network Science and Engineering, and IEEE Transactions on Vehicular Technology. He received the IEEE CSIM Technical Committee Best Journal Paper Award in 2018, IEEE ComSoc Young Author Best Paper Award in 2017, and IEEE ComSoc Asia-Pacific Best Young Researcher Award in 2019.
\end{IEEEbiography}
\begin{IEEEbiography}[{\includegraphics[width=1in,height=1.25in,clip,keepaspectratio]{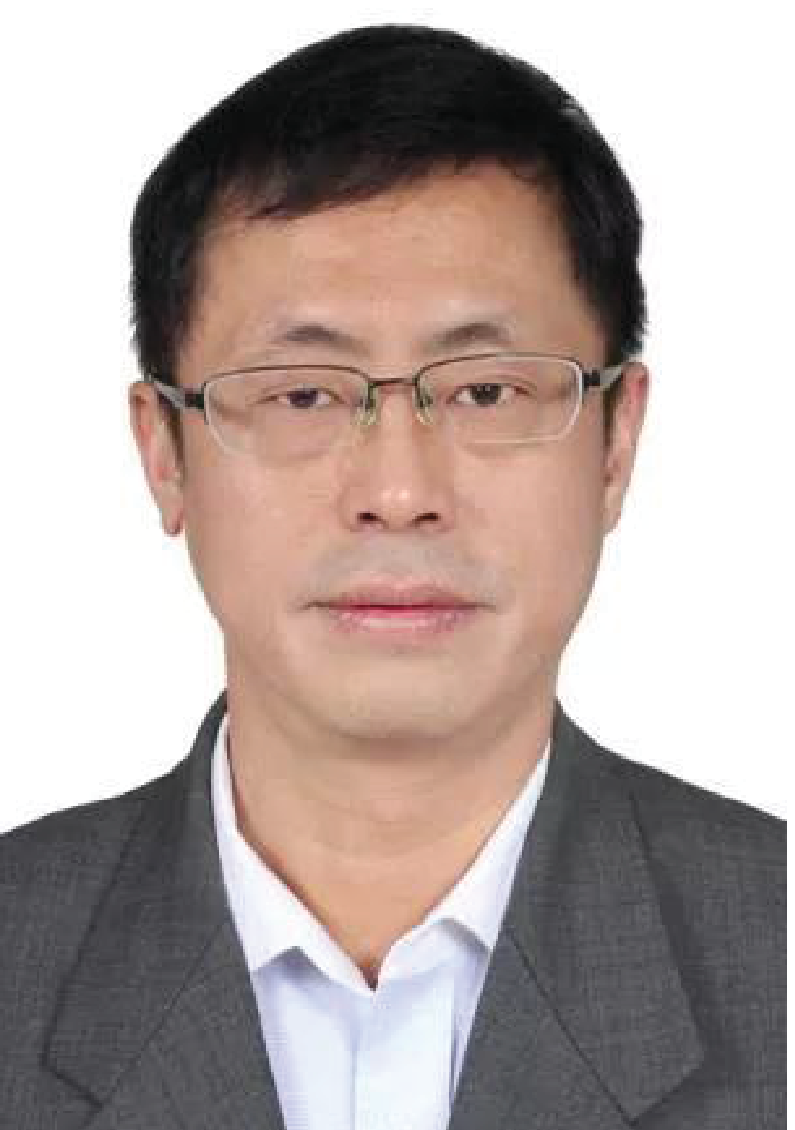}}]{Keping Long}
	(SM’06) received the M.S. and Ph.D. degrees from the University of Electronic Science
	and Technology of China, Chengdu, in 1995 and 1998, respectively. From September 1998 to August 2000, he was a Postdoctoral Research Fellow at the National Laboratory of Switching Technology and Telecommunication Networks, Beijing University of Posts and Telecommunications (BUPT),
	China. From September 2000 to June 2001, he was an Associate Professor at BUPT. From July
	2001 to November 2002, he was a Research Fellow with the ARC Special Research Centre for Ultra Broadband Information Networks (CUBIN), University of Melbourne, Australia. He is currently a
	professor and Dean at the School of Computer and Communication Engineering, University of Science and Technology Beijing. He has published more than 200 papers, 20 keynote speeches, and invited talks at international and local conferences. His research interests are optical Internet technology, new generation network technology, wireless information networks, value added services, and secure technology of networks. Dr. Long has been aTPC or ISC member of COIN 2003/04/05/06/07/08/09/10, IEEE IWCN2010, ICON2004/06, APOC2004/06/08, Co-Chair of the organization Committee for IWCMC2006, TPC Chair of COIN 2005/08, and TPC Co-Chair of COIN 2008/10. He was awarded by the National Science Fund for Distinguished Young Scholars of China in 2007 and selected as the Chang Jiang Scholars Program Professor of China in 2008. He is a member of the Editorial Committees of Sciences in China Series F and China Communications.
\end{IEEEbiography}

\begin{IEEEbiography}[{\includegraphics[width=1in,height=1.25in,clip,keepaspectratio]{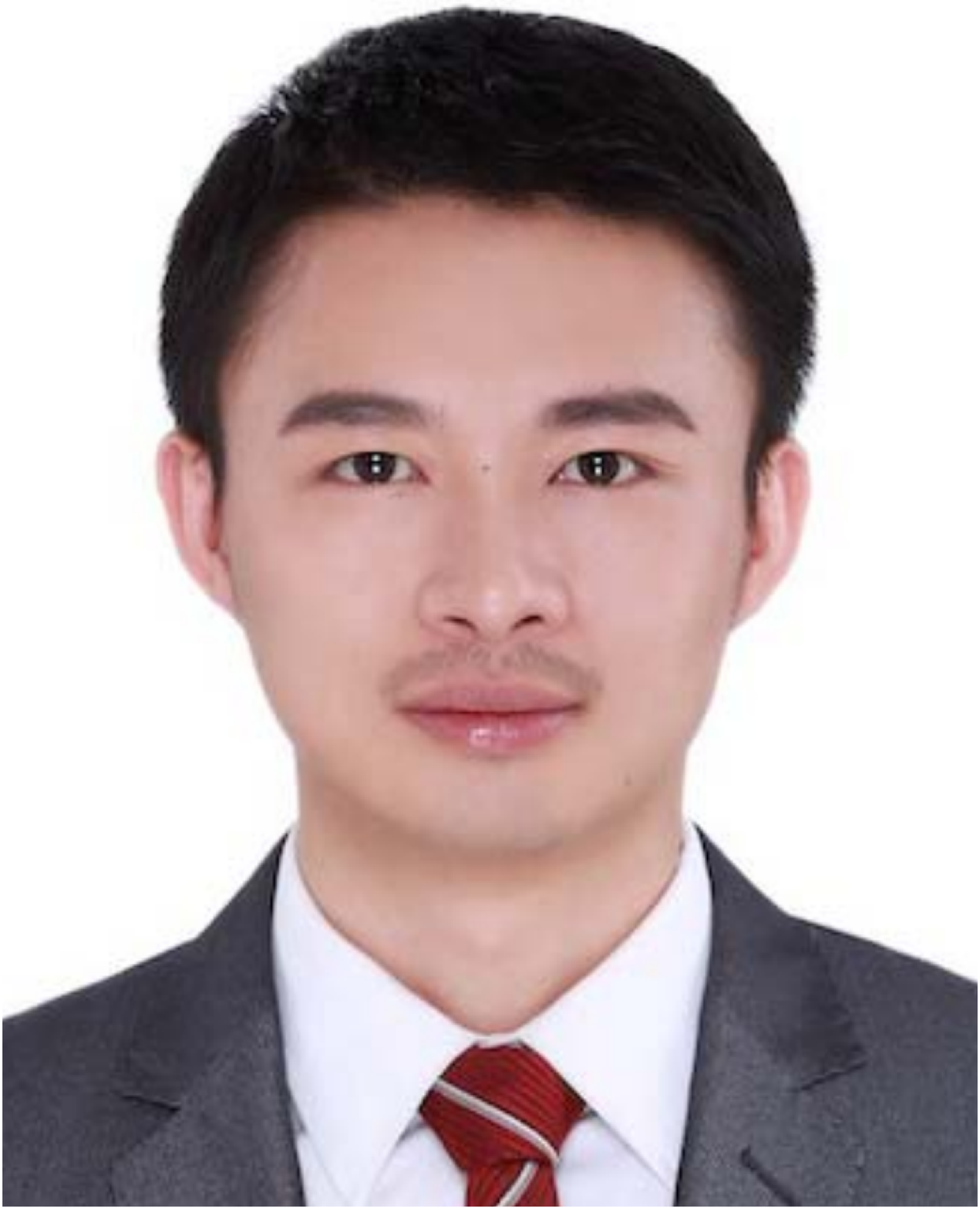}}]{Mingyu Zhou} received the Ph.D. degree from the Beijing University of Posts and Telecommunications (BUPT) as a focus on research into key technologies of wireless communications. After graduation in 2008, he became a Senior Engineer in Huawei Technologies, dedicated in 3GPP standardization and patent application. In 2014, he joined Baicells Technologies as the Research Director. Until now, he has applied for more than 100 patents (tens of them are PCTs), published more than 20 academic papers, and finished more than 100 standardization proposals. He was qualified in the Beijing Nova Program and has participated in several science and technology projects, targeting to bring more innovation to the 5G system.
\end{IEEEbiography}
\begin{IEEEbiography}[{\includegraphics[width=1in,height=1.25in,clip,keepaspectratio]{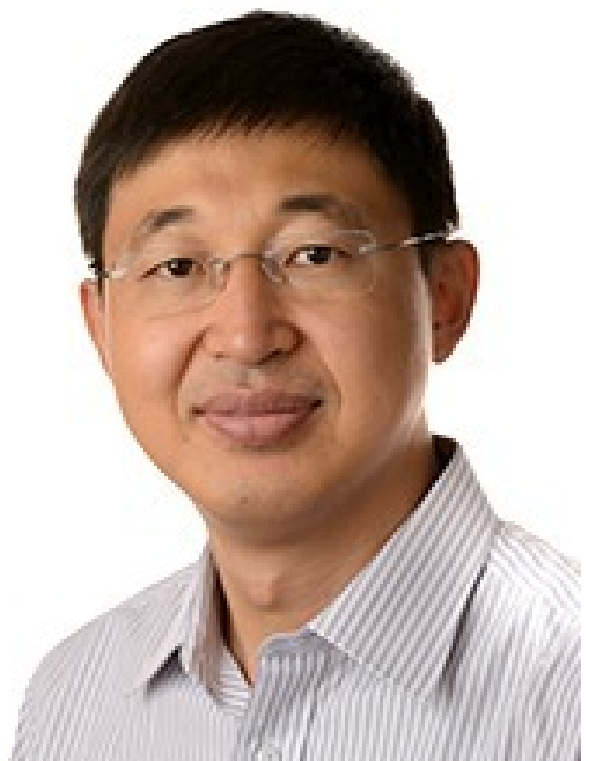}}]{Yonghui Li}(M’04, SM’09, F’19) received his PhD degree in November 2002 from Beijing University of Aeronautics and Astronautics. Since 2003, he has been with the Centre of Excellence in Telecommunications, the University of Sydney, Australia. He is now a Professor and Director of Wireless Engineering Laboratory in School of Electrical and Information Engineering, University of Sydney. He is the recipient of the Australian Queen Elizabeth II Fellowship in 2008 and the Australian Future Fellowship in 2012. He is a Fellow of IEEE. 
His current research interests are in the area of wireless communications, with a particular focus on MIMO, millimeter wave communications, machine to machine communications, coding techniques and cooperative communications. He holds a number of patents granted and pending in these fields. He is now an editor for IEEE transactions on communications, IEEE transactions on vehicular technology. He also served as the guest editor for several IEEE journals, such as IEEE JSAC, IEEE Communications Magazine, IEEE IoT journal, IEEE Access. He received the best paper awards from IEEE International Conference on Communications (ICC) 2014, IEEE PIRMC 2017 and IEEE Wireless Days Conferences (WD) 2014.
\end{IEEEbiography}

\begin{IEEEbiography}[{\includegraphics[width=1in,height=1.25in,clip,keepaspectratio]{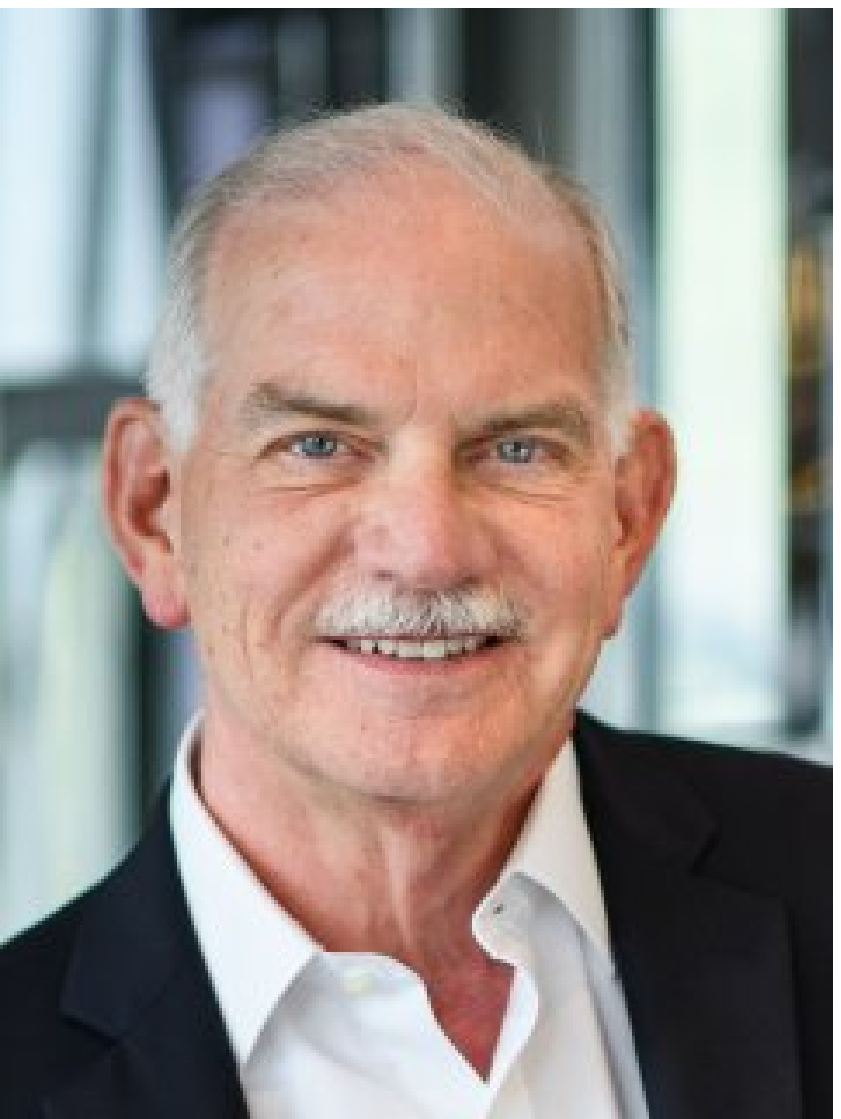}}]{H. Vincent Poor}
(S’72, M’77, SM’82, F’87) received the Ph.D. degree in EECS from Princeton University in 1977.  From 1977 until 1990, he was on the faculty of the University of Illinois at Urbana-Champaign. Since 1990 he has been on the faculty at Princeton, where he is currently the Michael Henry Strater University Professor. During 2006 to 2016, he served as the dean of Princeton’s School of Engineering and Applied Science. He has also held visiting appointments at several other universities, including most recently at Berkeley and Cambridge. His research interests are in the areas of information theory, machine learning and network science, and their applications in wireless networks, energy systems and related fields. Among his publications in these areas is the forthcoming book Machine Learning and Wireless Communications.  (Cambridge University Press). Dr. Poor is a member of the National Academy of Engineering and the National Academy of Sciences and is a foreign member of the Chinese Academy of Sciences, the Royal Society, and other national and international academies. He received the IEEE Alexander Graham Bell Medal in 2017.
\end{IEEEbiography}

\vfill

\end{document}